\documentclass[twocolumn]{aastex702}

\usepackage{amsmath}
\usepackage{amssymb}
\usepackage{CJKutf8}
\usepackage{xurl}
\usepackage{bm}

\newcommand{\tzol}{Thorne–\.Zytkow objects}
\newcommand{\tzo}{T\.ZO}
\newcommand{\tzos}{T\.ZOs}

\newcommand{\code}[1]{\texttt{#1}}
\newcommand{\mesa}{\code{MESA}}

\newcommand{\be}{\begin{equation}}
\newcommand{\ee}{\end{equation}}
\newcommand{\e}{\varepsilon}

\def \d {{\rm d}}

\begin{document}
\begin{CJK*}{UTF8}{gbsn}

\title{Neutron Star–Main-Sequence Star Collision:\\ Formation and Long-Term Evolution of Thorne–\.Zytkow Objects}

\author[orcid=0009-0000-8506-8503]{Yuhao Guo}
\affiliation{Tsung-Dao Lee Institute, Shanghai Jiao-Tong University, 1 Lisuo Road, Shanghai 201210, People’s Republic of China} 
\email[show]{yuhaoguo@sjtu.edu.cn}

\author[orcid=0000-0001-9565-9462]{Xiao-Tian Xu ({\upshape 徐啸天})}
\affiliation{Tsung-Dao Lee Institute, Shanghai Jiao-Tong University, 1 Lisuo Road, Shanghai 201210, People’s Republic of China}
\email[show]{xxu-tdli@sjtu.edu.cn}

\author[orcid=0000-0002-1934-6250]{Dong Lai}
\affiliation{Tsung-Dao Lee Institute, Shanghai Jiao-Tong University, 1 Lisuo Road, Shanghai 201210, People’s Republic of China}
\affiliation{Center for Astrophysics and Planetary Science, Department of Astronomy, Cornell University, Ithaca, NY 14853, USA}
\email[show]{donglai@sjtu.edu.cn}
\makeatletter
\def\@@email{yuhaoguo@sjtu.edu.cn\\xxu-tdli@sjtu.edu.cn\\donglai@sjtu.edu.cn}
\makeatother

\begin{abstract}
\tzol~(\tzos) are stars containing a neutron-star (NS) or black-hole (BH) core embedded in a stellar envelope and powered primarily by mass accretion onto the core. 
Classical evolution models of \tzos\  begin with an assumed equilibrium configuration and predict chemical anomalies in the stellar envelope resulting from nuclear burning in the vicinity of the NS core. Here we connect dynamical formation to long-term evolution by combining a three-dimensional simulation of a post-supernova collision between a $1.4M_\odot$ NS and a $10M_\odot$ main-sequence star with a subsequent stellar evolution calculation using MESA. The collision successfully produces a bound, quasi-spherical \tzo\ with the NS settled at the center of the stellar envelope. In contrast to the classical picture, the collision heats a substantial fraction of the star and leads to widespread nuclear burning during the subsequent thermal-relaxation phase. As the envelope becomes convective, material processed by the carbon--nitrogen cycle is transported to the surface, leaving a potentially observable chemical imprint of formation that persists after the star attains the standard accretion-supported \tzo\ structure. At the end of our simulation, the stellar luminosity greatly exceeds Eddington in the outer envelope, suggesting that significant mass ejection will occur, leaving behind an isolated NS or BH.
\end{abstract}

\keywords{\uat{High Energy astrophysics}{739} --- \uat{Stellar astronomy}{1583}}

\section{Introduction} 

\tzol \, (\tzos) are hypothetical stars consisting of a neutron-star (NS) or black-hole (BH) core embedded in a stellar envelope \citep{1975ApJ...199L..19T,1977ApJ...212..832T}, whose
structure is supported by either the accretion power of the NS or nuclear burning \citep{1991ApJ...380..167B}. 
Observationally, massive \tzos\, are expected to have unique chemical signatures at the surface because of the rapid proton capture process at the base of the convective envelope
\citep{1993MNRAS.263..817C,1995MNRAS.274..485P}.
Accordingly, some evolved stars with peculiar chemical abundance have been proposed to be the candidates of \tzos\, \citep[e.g., HV2112;][]{2014MNRAS.445L..36T}, whose nature remains under debate \citep{2018MNRAS.479.3101B,2023MNRAS.524.1692F}. 
Identifying a T\.ZO from its surface properties relies on predictions concerning whether such an object can form and how it subsequently evolves. Recent studies have questioned whether the merger of an NS and an MS star can produce the classical T\.ZO  \citep{2024ApJ...971..132E, 2024ApJ...977..196H, 2026arXiv260423503C, 2026arXiv260809395C}, and evolutionary calculations find that the surface properties of T\.ZOs may differ from the classical expectations \citep{2023MNRAS.524.1692F}. Thus, establishing robust observational signatures requires models that connect the formation of a T\.ZO and its long-term evolution.

Many scenarios for the formation of \tzos\, have been proposed \citep[see][for a review]{2026enap....3..336O}, 
one of which involves supernova (SN) kicks in close binaries \citep{1994ApJ...423L..19L}. The asymmetries in SN ejecta generate a momentum kick on the newborn NS, widening the orbit and pumping eccentricity if the binary is not disrupted.
Meanwhile, the SN ejecta may heat up the companion star, causing envelope inflation  \citep{Liu:2015tya,Hirai:2025oyi, Cary2026ApJ...996..141C}. 
With these two effects, it is possible that the pericenter separation of the post-SN orbit is small enough that a direct collision between the NS and the companion occurs \citep{1994ApJ...423L..19L}, creating the possibility for the NS to settle down to the center of the companion star. 
Previous works have suggested that such a collision can indeed lead to the formation of dynamically stable \tzos\ \citep[][see also Y. Shi et al. \citeyear{Shi:2026kci}]{2022MNRAS.517.4544H, 2025ApJ...993...61W}. 

However, rapid accretion of the NS within the stellar envelope may trigger NS collapse, preventing the formation of a long-lived \tzo\ \citep{2026arXiv260809395C, 2026arXiv260423503C}. The accretion rate within the envelope is usually expected to be Eddington-limited \citep{Ivanova2013}, but efficient neutrino cooling can enable hypercritical accretion 
\citep{1993ApJ...411L..33C,CHEVALIER199595,1996ApJ...460..801F,1996ApJ...459..322C,2012ApJ...752L...2C,Anninos:2025fti}. On the other hand, angular momentum can limit the accretion rate \citep{2015ApJ...798L..19M,2015ApJ...803...41M}, potentially allowing a long-lived \tzo\ to form.

Independently, 
the basic structure of \tzos\ was established by assuming hydrostatic and thermal equilibrium \citep{1975ApJ...199L..19T,1977ApJ...212..832T,1991ApJ...380..167B}.
We refer to this equilibrium configuration as the standard \tzo\, structure.
Later studies relied on these equilibrium configurations as starting models to explore the subsequent long-term evolution, nucleosynthesis signatures, and provide further predictions for identifying \tzos\ in observations \citep{1992ApJ...386..206C,1993MNRAS.263..817C,1995MNRAS.274..485P, 2023MNRAS.524.1692F} and the final fate of \tzos\ \citep{1995MNRAS.274..485P, 2026enap....3..336O}. 

Nevertheless, studies that follow the evolution of T\.ZOs from the post-SN orbit, hydrodynamical interactions, to the long-term evolution are still missing.  
As shown by \citet{2025ApJ...993...61W}, after the hydrodynamical interactions the stellar structure can deviate considerably from single stars.
In particular, the core temperature is much higher than the typical temperature of the CNO cycle.
So far, it is unclear whether and how this structure evolves to thermal equilibrium; this thermal relaxation process might affect the long-term evolution, and even the final fate of a \tzo\,. 
We aim to bridge the dynamical formation of a \tzo\ triggered by SN kick and NS-MS collision, and the long-term evolution to investigate how the stellar structure after the formation adjusts towards the standard \tzo\, structure. In this work, we adopt an MS stellar model (unaffected by SN ejecta) as a reference framework to study the impact of the formation process on the long-term evolution, while a broader exploration of the parameter space and more realistic simulations are left for future studies.  

This paper is structured as follows. In Section~\ref{sec:meth} we present our methodology of 3D hydrodynamic simulation, 1D stellar evolution and numerical setup of our reference model. In Section~\ref{sec: result}, we present our results. 
Finally, we discuss several issues related to our results in Section  \ref{sec: discuss} and conclude in Section~\ref{sec: conc}.

\section{Method}
\label{sec:meth}

\begin{figure}[!tp]
    \centering
    \includegraphics[trim=0 0.0cm 0.0cm 0cm,clip,width=0.9\linewidth]{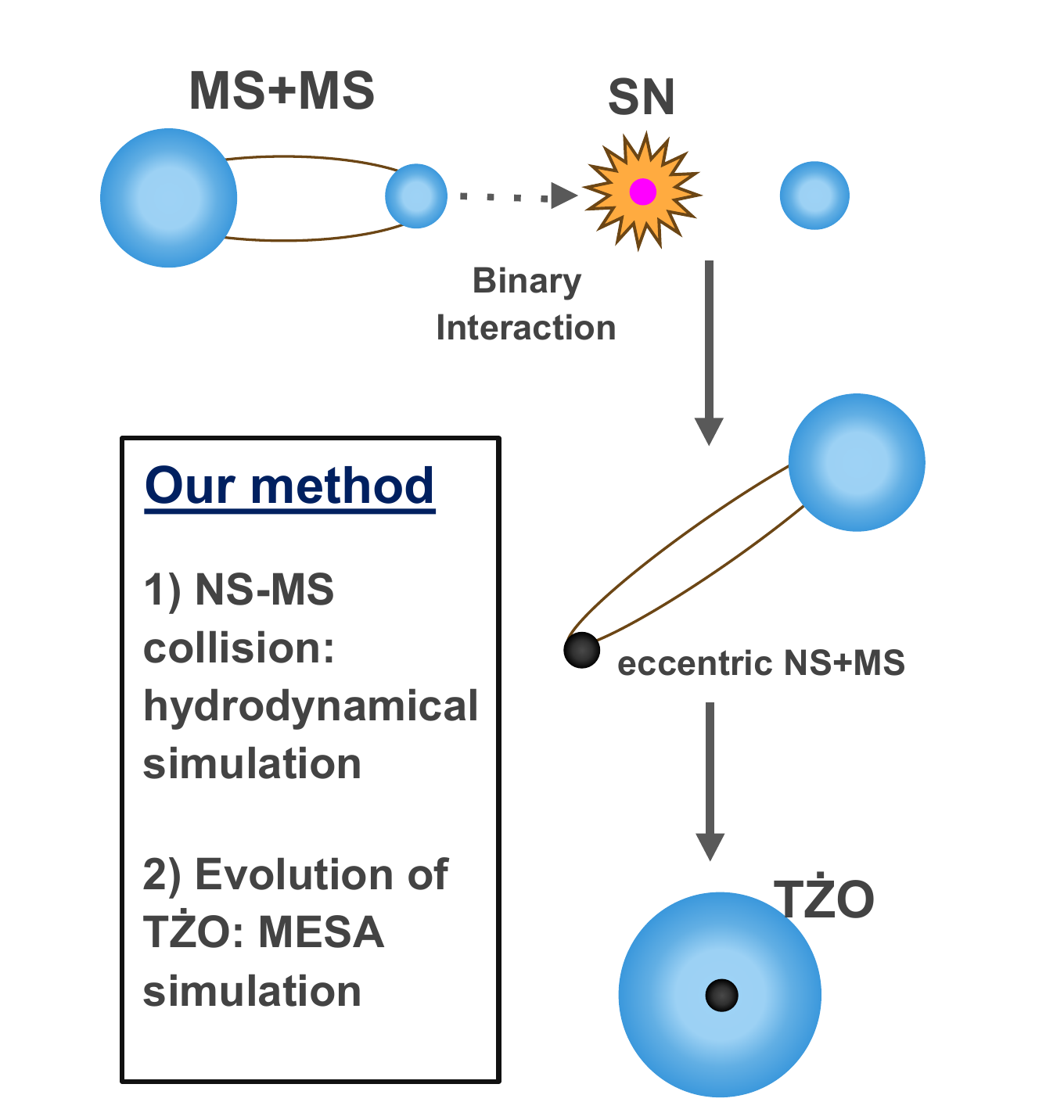}
    \caption{
    The scope and method of this work. We envision a supernova (SN) explosion in an isolated binary. The prior binary interaction is not considered. The post-SN orbit is highly eccentric such that the newborn NS collides with the main-sequence (MS) companion. This process is modeled through a 3D hydrodynamical simulation. After the dynamical assembly of the \tzo, we use the 1D stellar evolution code MESA to study the long-term evolution (See Section \ref{sec:meth} for details). 
    }
    \label{fig:schematic-TZO}
\end{figure}

We present a schematic to illustrate the scope of our work in Fig.\,\ref{fig:schematic-TZO}.
We consider a \tzo\ formed from the collision between an NS and an MS star. 
The companion is expected to be an MS star because the NS progenitor should be initially more massive and have a shorter lifetime than that of the secondary star. Consequently, when the primary star reaches core collapse, the secondary star is still on the MS.  
For simplicity, we use a zero-age main-sequence (ZAMS) model for the MS star, and our simulation starts from the post-SN eccentric orbits, assuming a suitable kick velocity and direction. We neglect the evolution of the MS star and the effects of the supernova ejecta on it, which we do not expect to significantly alter the main conclusions of this work (cf., Sect.\,\ref{sec: discuss}). 

We choose our reference model to be a $1.4 M_{\odot}$ newborn NS and a $10 M_{\odot}$ ZAMS. The choice of $10 M_{\odot}$ is based on past population synthesis predictions on Be X-ray binaries \citep[e.g.,][]{Shao2014ApJ...796...37S,Xu2025A&A...704A.218X,Schurmann2025arXiv250323878S}.
For the post-SN orbital configuration, 
we adopt a semi-major axis of $a = 9 R_{\odot}$ and an eccentricity of $e = 0.8$ for the post-SN orbit\footnote{We test the combinations of $a$ and $e$ for $a=8,9,10 R_{\odot}$ and $e=0.7,0.8,0.9$. The change of the initial orbital configuration does not have considerable effects on our results.}, corresponding to an initial orbital period $P_{\rm orb,i}\approx\mathrm{1 \,d}$ 
and a pericenter distance of $r_{\mathrm{p}} = 1.8R_{\odot}$, which is smaller than our initial companion star size $R_{\ast}^{\rm ini}\approx 4 R_{\odot}$, allowing for the NS--MS collision. 

Following the method illustrated in Fig.\,\ref{fig:schematic-TZO},
we perform a 3D hydrodynamical simulation for the NS--MS collision by using  \texttt{ENZO} \citep{Brummel-Smith2019}, where we generate a non-rotating ZAMS model for the companion star by using the 1D stellar evolution code \mesa\,r24.08.1 \citep[Modules for Experiments in Stellar Astrophysics;][]{paxton:11, paxton:13, paxton:15, paxton:18, paxton:19, jermyn22}.
After the collision assembles a dynamically stable \tzo, we
map the configuration into the MESA code for long-term evolution. 
In Sect.\,\ref{meth: 3D hydro}, we outline our numerical setup for 3D hydrodynamical simulation. In Sect.\,\ref{meth: 3D-1D}, we describe the method on 3D-1D mappings. In Sect.\,\ref{meth: 1D stellar}, we give our setup for 1D stellar evolution.

\subsection{NS--MS collision: 3D Hydrodynamical Simulation}
\label{meth: 3D hydro}

The \texttt{ENZO} code solves the Eulerian hydrodynamic equations, and the self-gravity of gas is computed by the Poisson solver in \texttt{ENZO}. We adopt the ideal gas equation of state (e.o.s) with an adiabatic index of $\gamma = 5/3$. We neglect nuclear energy release during the hydrodynamical simulation, assuming that its effect on the dynamics is small. We also neglect chemical abundance evolution and assume a fixed composition (for each parcel of  gas) during this phase.
We evolve the NS and the MS star (i.e., the self gravitating gas) simultaneously within a grid described in Sect.\,\ref{sec: grid}. The NS is modeled as a Newtonian point source with a softening radius (i.e. the effective size of the NS $R_{\rm core}$) set to $0.05R_{\odot}$; it produces energy  at the Eddington luminosity, which heats up the gas that the NS passes through (Sect.\,\ref{sec: NS}).  
We assume the accretion of the NS is limited by the Eddington rate (see below). Since the timescale of the NS-MS collision is too short for the NS to accrete a considerable amount of material, we can neglect the mass change of the NS.

\subsubsection{Simulation grid}
\label{sec: grid}

The computational domain is a Cartesian box centered on the binary center of mass (c.o.m.) with a half-size of $51.2R_{\odot}$. The base grid has a resolution of $0.4R_{\odot}$, with a finest resolution of $0.0125R_{\odot}$. 
To resolve the main interaction region during pericenter passage, we employ three nested fixed refinement boxes centered on the domain origin, with half-sizes of $9$, $9.5$, and $10R_{\odot}$. Together, they ensure that the global merger structure is resolved with a resolution of $0.05R_{\odot}$.

Additionally, since the NS is initially outside the fixed refinement region and the final merger product can acquire a recoil velocity due to asymmetric mass ejection during the NS-MS collision,
we adopt adaptive mesh refinement following the NS, 
ensuring the majority of the mass of the resulting \tzo\, stays at the center of the refined computational domain. 
To resolve both the global gas structure of the \tzo\, and the local gravitational field of the NS, we introduce five nested refinement regions centered on the instantaneous NS position, with radii of $4.0\,R_\odot$, $3.5\,R_\odot$, $3.0\,R_\odot$, $0.5\,R_\odot$ and $0.3\,R_\odot$. The three outer levels cover the region of the \tzo\, and maintain a resolution comparable to that of the central fixed-refinement grid. The two innermost levels resolve the effective numerical radius of the NS, $R_{\rm core}$, with four grid cells, so that the softened-gravity region is sampled by approximately $8^3=512$ cells. 

\subsubsection{
Neutron star in hydrodynamical simulation
}
\label{sec: NS}

The gravitational potential of the NS is softened within an effective radius of NS as $R_{\rm core}$,
using the cubic spline kernel introduced in \citet{2007MNRAS.374.1347P}. We note that the adopted $R_{\rm core}$ is much larger than the radius of a realistic NS due to the resolution limit of our 3D simulation.
We assume the NS isotropically radiates at the Eddington luminosity,
\be
\label{eq: Ledd}
L_{\mathrm{Edd}} = \frac{4\pi c G M_{\mathrm{NS}}}{\kappa},
\ee
where $G$ is the gravitational constant, $c$ is the speed of light, and 
$\kappa=0.2\, (1+X)\,\mathrm{cm}^2\, \mathrm{g}^{-1}$ is the Thomson opacity for fully ionized hydrogen and helium with $X$ the hydrogen mass fraction.
We take $\kappa=0.34\,\mathrm{cm}^2\, \mathrm{g}^{-1}$ (i.e., $X=0.7$), corresponding to $L_{\mathrm{Edd}}\approx5\times10^4L_\odot$.
We further assume that a fraction, $f_{\rm th}$, of the radiation energy is converted into local thermal energy at a rate $L_{\rm th}$.  
Such heating should be more efficient in optically thicker regions (i.e., higher density), and we therefore introduce the following prescription,
\be
\frac{L_{\rm th}}{L_{\rm Edd}}\equiv f_{\rm th} = \exp\left(-\rho_0/\rho_{\rm loc}\right),
\ee 
where $\rho_0$ is a characteristic density scale separating optically thin and thick regimes, and $\rho_{\mathrm{loc}}\equiv \rho(\bm{r}_{\rm NS})$ is the local density at the NS position.
This function ensures a smooth transition from inefficient coupling ($f_{\rm th} \rightarrow 0$ for $\rho_{\rm loc} \ll \rho_0$) to full thermalization ($f_{\rm th} \rightarrow 1$ for $\rho_{\rm loc} \gg \rho_0$). 
In the following, we describe the modified energy equation with the effects of heating. 

The thermal energy converted from radiation is then deposited into the surrounding gas within a finite spherical region of a volume $V_{\mathrm{th}} \equiv 4/3  \pi r_{\mathrm{th}}^3$, which is introduced to avoid numerical singularities associated with point-like energy injection.
The deposited energy is spatially distributed using a normalized smoothing kernel $S(\bm{r}, r_{\mathrm{th}})$ satisfying $\int_{|\bm{r}-\bm{r}_{\mathrm{NS}}|<r_{\mathrm{th}}} S(\bm{r}, r_{\mathrm{th}})\, d^3r = V_{\mathrm{th}}.$ 
With this prescription, the gas energy equation is modified by an additional source term,
\be
\frac{\partial \rho \e}{\partial t}
+ \nabla \cdot [(\rho \e + P)\bm{v}]
= -\rho \bm{v} \cdot \nabla \Phi
+ \frac{L_{\mathrm{th}}}{V_{\mathrm{th}}} S(\bm{r}, r_{\mathrm{th}}),
\ee
where $\rho$, $\bm{v}$, and $P$ denote the gas density, velocity, and pressure, respectively, and $\e \equiv u + |\bm{v}|^2/2$ is the specific total energy  with $u$ being specific internal energy. The gravitational potential $\Phi$ includes contributions from both the self-gravity of the gas and the softened NS potential.

The injection scale is chosen as $r_{\mathrm{th}} = 0.1 R_{\odot}$, which is larger than the gravitational softening radius, ensuring that the energy deposition is numerically resolved and spatially smooth. The typical density scale is chosen as $\rho_0 = 10^{-6} \text{g}/\text{cm}^{3}$. 
We choose the typical density scale such that the heating happens within the stellar envelope. The density scale should then satisfy the condition that it is much larger than the minimum density limit of the stellar profile $10^{-9} \text{g}/\text{cm}^{3}$ and much lower than the mean density $1 \text{g}/\text{cm}^{3}$. However, we test different $r_{\rm th}$ and $\rho_0$ and we find different values do not significantly change the profile of the final object.

\subsubsection{
Initial model for the main-sequence companion
}
\label{sec: MS}

We use the MESA code to generate a $10\,M_{\odot}$ non-rotating ZAMS star with a metallicity of $Z=0.02$ for the companion star. The Eulerian hydrodynamical variables, initial density, velocity, and internal energy fields $\rho_{\rm ini}(\bm{r}), \bm{v}_{\rm ini}(\bm{r}), u_{\rm ini}(\bm{r})$ are assigned in a three dimensional grid based on a one dimensional solved stellar model. As our simulation is conducted in the c.o.m. frame, the MS initially is located at the apocenter of the orbit, and the initial velocity $\bm{v}_{\rm ini}(\bm{r})$ is set to be the Keplerian orbital velocity.

Additionally, since the Eulerian hydrodynamics solver in \texttt{ENZO} cannot handle vacuum, we add a low-density, high-temperature non-moving ambient medium outside the star. The medium is dilute enough not to affect our results.

\subsection{1D stellar profile inferred from 3D structure}
\label{meth: 3D-1D}

We evolve the system using 3D hydrodynamics until the NS is well settled at the center of the star and total bound gas mass converges to a nearly constant value, which we adopt as the criterion that the remnant has reached a quasi-steady configuration (see Eq.\,\ref{eq:Mast} below). We then map the 3D simulation output to 1D radial profiles for the subsequent stellar evolution by using the approach described in  
\citet{2016MNRAS.463.3461S}.

For thermodynamic quantities, we perform angle-averaging on spherical shells in a spherical coordinate system $(r, \theta, \varphi)$ centered at the NS\footnote{Note that this is different from the simulation coordinate system in the c.o.m. frame}  with the orbital plane as its equatorial plane. 
For a given radius $r$, we calculate the angle-averaged quantities as
\be
\label{eq: average}
\begin{split}
\rho_{\ast}(r) &= \frac{1}{4\pi}\int_{\sigma(r)} \rho(r, \theta, \varphi)\, \d\Omega, \quad \\
u_{\ast}(r) &= \frac{1}{4\pi}\int_{\sigma(r)} u(r, \theta, \varphi)\, \d\Omega, 
\end{split}
\ee
where $\d\Omega \equiv \sin\theta\, \d\theta\, \d\varphi$ is the solid angle, and $\sigma(r)$ denotes the spherical surface at radius $r$. The ``$*$" subscript  denotes angle-averaged quantities. For the angular momentum profile, we compute a mass-weighted average specific angular momentum over a thin spherical shell with width $\Delta r$:
\be
j_{\ast}(r) = \frac{\int_{r}^{r+\Delta r} \rho(\bm{r}) \left[\left(\bm{r}\times \bm{v}\right)\cdot \hat{\bm{z}}\right] \d^3 \bm{r}}{\int_{r}^{r+\Delta r} \rho(\bm{r})\, \d^3 \bm{r}}.
\label{eq:avg-j}
\ee
The lower and upper bounds of the radius of these 1D profiles are $R_{\rm core}$ and stellar radius $R_\ast$ respectively, where $R_\ast$ is defined as the boundary between the stellar material and the unbound gas. We estimate the mass of the bound gas $M_{\ast}$ through the Bernoulli criterion, which is
\be
\begin{split}
M_{\rm \ast}&=
\int_{B<0}
\rho\,\d^3\bm{r}, \\
B &\equiv \frac{1}{2} \rho \bm{v}^2 + \rho \Phi + \rho u + P .
\label{eq:Mast}
\end{split}
\ee
We then define the corresponding stellar surface $R_{\ast}$ with 
\be
\label{eq:Rast}
M_{\ast}
=
\int_{0}^{R_{\ast}}
\rho_{\ast}(r)\,4\pi r^2\d r.
\ee
This Bernoulli criterion provides an approximate estimate of the bound material, although it may not exactly represent the final stellar mass because the Bernoulli constant is not conserved given the presence of shock.
We vary the adopted value of $M_{\ast}$ by up to $5\%$ with corresponding $R_{\ast}$ and find that the resulting one-dimensional \mesa\, models are insensitive to this choice.

\subsection{Long-term evolution: 1D Stellar evolution}

Our MESA calculation is initialized by an NS core with an effective core radius. For the stellar material above the NS core, we use the profiles mapped from the 3D hydrodynamical simulation. We describe our approach of generating the initial model for the MESA calculations in Sect.\,\ref{sec: core} and the input physics for MESA in Sect.\,\ref{sec: stellar evo}. 

\subsubsection{Initial model of MESA}
\label{sec: core}

Consistent with the 3D simulation,
the mass of the NS core $M_{\rm core}$ is initially taken to be $1.4\,M_\odot$, the effective core radius is equal to $R_{\rm core}$ (i.e., 0.05$R_{\odot}$), and the accretion luminosity of the NS core is taken to be the Eddington luminosity (Eq.~\ref{eq: Ledd}), while the opacity uses the default opacity table in \mesa. 
The mass growth of the core is governed by the Eddington accretion rate, 
\be
\dot{M} = \frac{L_{\mathrm{Edd}}}{\eta c^2}\approx 2.6\times 10^{-8}\left( \frac{M_{\rm core}}{M_\odot}\right)\left(\frac{0.1}{\eta}\right){\rm M_{\odot} yr}^{-1},
\label{eq:Mdot-edd}
\ee
where $\eta$ is the accretion efficiency. For the envelope above the NS core, we use the mapped profiles to describe its structure (cf., Sect.\,\ref{meth: 3D-1D}). Since we do not expect significant nuclear reaction during the hydrodynamical phase, the metallicity $Z$ is taken to be the same as the initial stellar model used in the hydrodynamical simulation (i.e., $Z=0.02$). 
The total mass of the system is given by $M_{\rm tot}=M_{\rm core}+M_\ast$, where $M_\ast$ is determined in the 3D simulation (cf., Sect.\,\ref{meth: 3D-1D}).

Numerically, we use the \texttt{relax\_core} method provided in \mesa\ to set up the NS core and the \texttt{relax\_initial\_entropy} method to ensure the structure of the initial model matches the mapped 1D profiles. In addition, the angular momentum profile is included through a similar method, \texttt{relax\_initial\_angular\_momentum}.

\subsubsection{Input physics of MESA}
\label{sec: stellar evo}

We adopt the same input physics as that in \citet{Wang2020ApJ...888L..12W} and \citet{Xu2025A&A...704A.218X}. In the following, we briefly describe the most relevant inputs and the differences from \citet{Xu2025A&A...704A.218X}. The boundary of convection zones is determined by the Ledoux criterion, which takes into account the effect of chemical gradient. We model convection by using the standard mixing-length theory with a parameter of $\alpha_{\rm MLT}=1.5$ \citep{Bohm-Vitense1958ZA.....46..108B}, and we consider the convective overshoot with a parameter of $\alpha_{\rm OV}=0.335$, as determined by \citet{Brott2011A&A...530A.115B}. While other types of mixing processes are also considered in our calculations, they are negligible in our reference model (cf., Fig.\,\ref{fig:kipp-diagram}). 

For the reference model, we include differential rotation. Following \citet{2023MNRAS.524.1692F}, we adopt the \texttt{approx21} nuclear network, which is suitable for the purpose of this work. We do not take into account the interrupted rapid-proton-capture process, as it occurs within $R_{\rm core}$, which is not resolved in our calculation, and it does not considerably contribute to the energy generation. For the same reason, the effects of general relativity are ignored. For simplicity, we turn off stellar wind for the reference model.  

In addition, in previous studies, models were terminated when the NS core becomes a black-hole (BH) core \citep[e.g.,][]{1992ApJ...386..206C,1993MNRAS.263..817C}, which occurs when the core mass reaches $2\,M_\odot$. However, our model evolves across this BH formation point, with the same treatment for the core radius ($R_{\rm core}$), accretion luminosity, and mass accretion rate, and we neglect the additional energy release at the BH formation point \citep[but see][]{2018MNRAS.475L..49M}. Then, we run the model until the MESA calculation stops, which we interpret as the breakdown of hydrostatic equilibrium. We discuss its implications in Sect.\,\ref{sec: discuss}.

\label{meth: 1D stellar}

\section{Result}
\label{sec: result}

\begin{figure*}[!tp]
    \centering
    \includegraphics[width=0.95\linewidth]{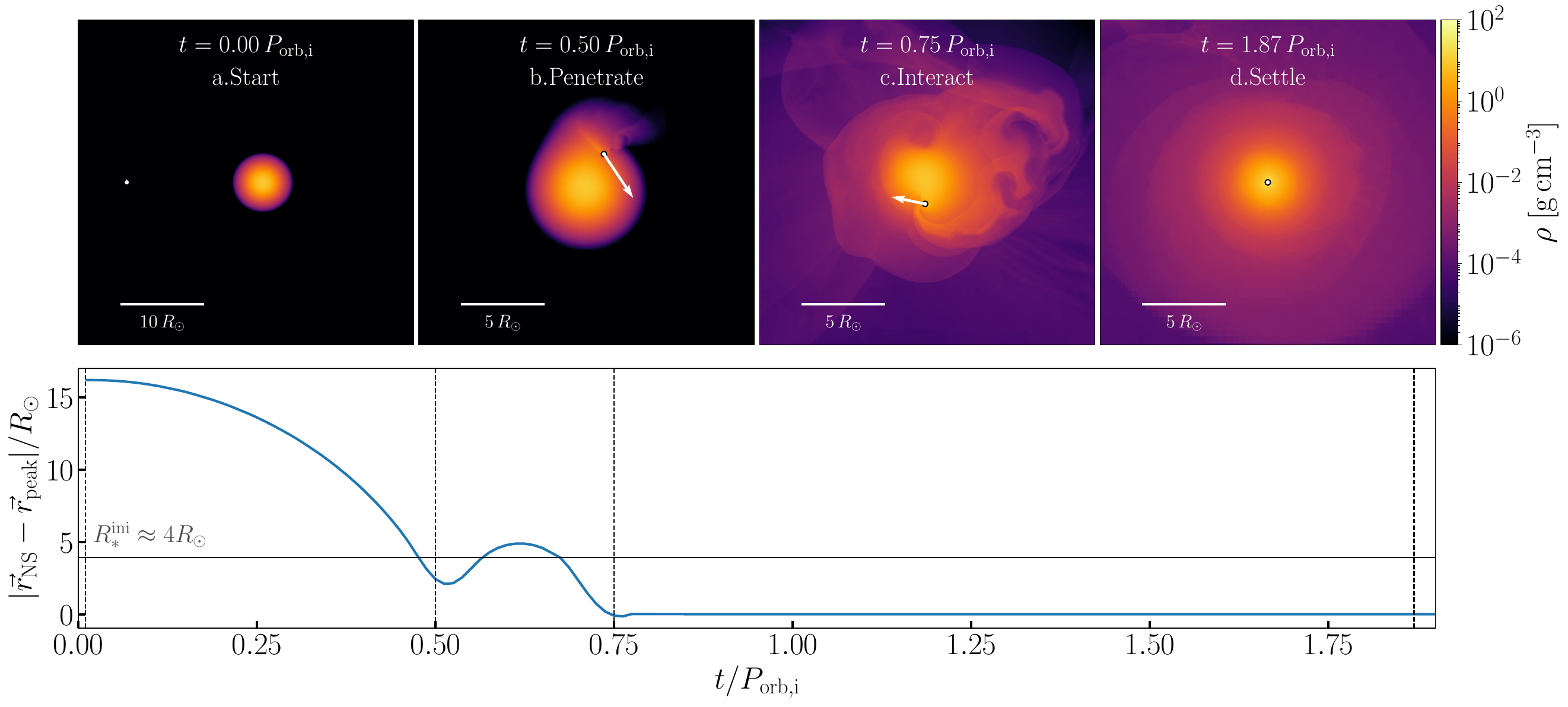}
    \caption{Evolution of a $1.4\,M_{\odot}$ NS interacting with a $10\,M_{\odot}$ companion star. The top panels show snapshots at different evolutionary stages of the system: (a) Start: the simulation begins with a post-SN NS--MS binary on a highly eccentric orbit; (b) Penetrate: the NS penetrates the envelope of the companion star; (c) Interact: strong dynamical interaction between the NS and the star disrupts the stellar envelope; (d) Settle: the system relaxes into a nearly spherical quasi-static configuration, forming a \tzo. The white circles indicate the position of the NS, while the arrow lengths are proportional to the velocity in the center-of-mass frame. Note that the initial and final velocity vectors are so small that they are nearly points.
    The evolutionary time ($t$) is shown in units of the initial orbital period above each panel. The bottom panel shows the separation between the NS and the peak-density location. Vertical dashed lines indicate the times corresponding to the four snapshots shown above. The horizontal line marks the initial radius of the main-sequence star $R_{\ast}^{\rm ini}$. }
    \label{fig:density-evolution}
\end{figure*}

\subsection{Dynamical process: 3D Hydrodynamical simulation}
\label{sec: 3D hydro}

\subsubsection{NS-MS collision}

We present the 3D hydrodynamical simulation of the NS-MS collision in Fig.~\ref{fig:density-evolution}.
The NS plunges into the MS star at the pericenter ($t=0.5P_{\rm orb,i}$),  
causing a strong shock near the stellar surface, which efficiently heats up and ejects the outer layers of the star. 
Meanwhile, orbital energy and angular momentum are dissipated by the gas friction,
resulting in a strong orbital shrinkage and the deceleration of the NS. 
After the first interaction, the maximum separation between the NS and peak density location is about $5 R_{\odot}$ (the lower panel of Fig.\,\ref{fig:density-evolution}), which is already comparable to the radius of the companion star before interaction (about $4 R_\odot$), meaning that the NS has already been embedded into the star.
The strong interaction then disrupts the outer part of the star at the second pericenter passage ($t=0.75P_{\rm orb,i}$). At the same time, the gas is focusing around the NS, making the NS position have the peak gas density (i.e., $|\bm{r}_{\rm NS}-\bm{r}_{\rm peak}|\approx 0$).
Finally, the gas falls back due to gravity.
From $t=1.87P_{\rm orb,i}$, 
$M_{\ast}$ reaches a constant value, and the stellar structure does not evolve considerably (i.e. not contracting or expanding). The entire structure relaxed to a spherical configuration. 
Using the criterion introduced in Sect.\ref{meth: 3D-1D}, we find $M_{\ast}/M_{\ast}^{\rm ini}\approx 90\%$, meaning that about $9M_{\odot}$ of stellar material is bound, giving a stellar radius of about $15\,R_\odot$.
Due to the asymmetric envelope ejection, the resulting \tzo\ has a small spatial velocity, and the remaining envelope material eventually co-moves with the NS. 
The corresponding stellar profile is taken to be the final profile given by the hydrodynamical simulation, which we present in the next subsection. 

\subsubsection{Final profile of hydrodynamical simulation}

\begin{figure*}[!tp]
    \centering
    \includegraphics[
    width=0.95\linewidth,
    height=0.65\textheight,
    keepaspectratio
    ]{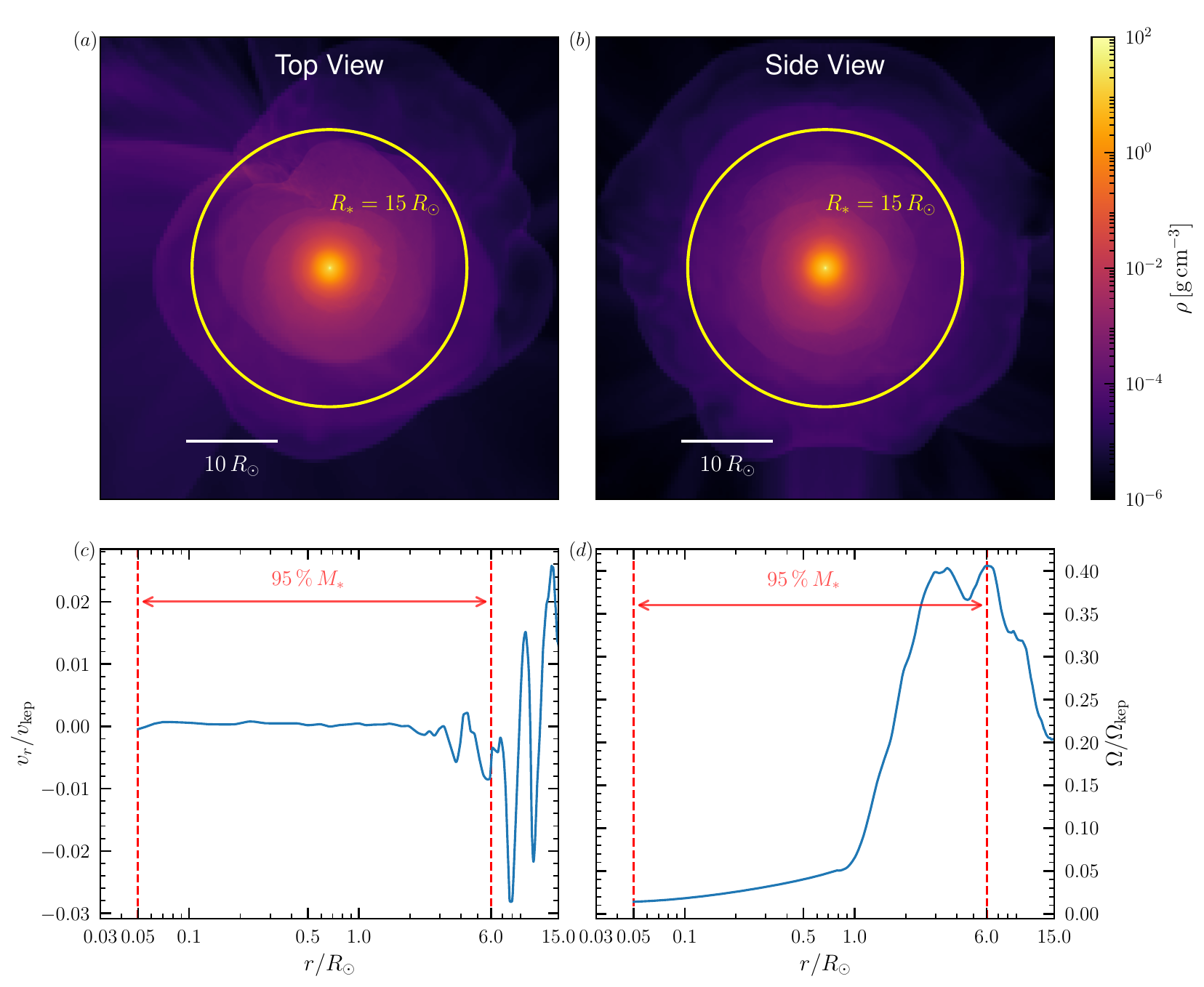}
    \caption{The ``settled'' \tzo\ configuration after the hydrodynamical phase. The top panels show density slices viewed from the orbital axis (top view; panel a) and perpendicular to the orbital plane (side view; panel b). The yellow circle indicates the spatial region adopted for the 3D-to-1D mapping procedure, and the white scale bar corresponds to $10R_{\odot}$. The bottom panels show the angle-averaged radial velocity (panel c) and angular velocity (panel d) profiles, normalized by the Keplerian velocity $v_{\rm Kep}$ and angular velocity $\Omega_{\rm Kep}$ at the equator, respectively. The red dashed lines mark the inner boundary of the stellar profile and the radius enclosing $95\%$ of the stellar mass, while the horizontal red arrows indicate the corresponding mass region. 
}
    \label{fig:final-configuration}
\end{figure*}

\begin{figure}[!tp]
    \centering
    \includegraphics[width=\linewidth]{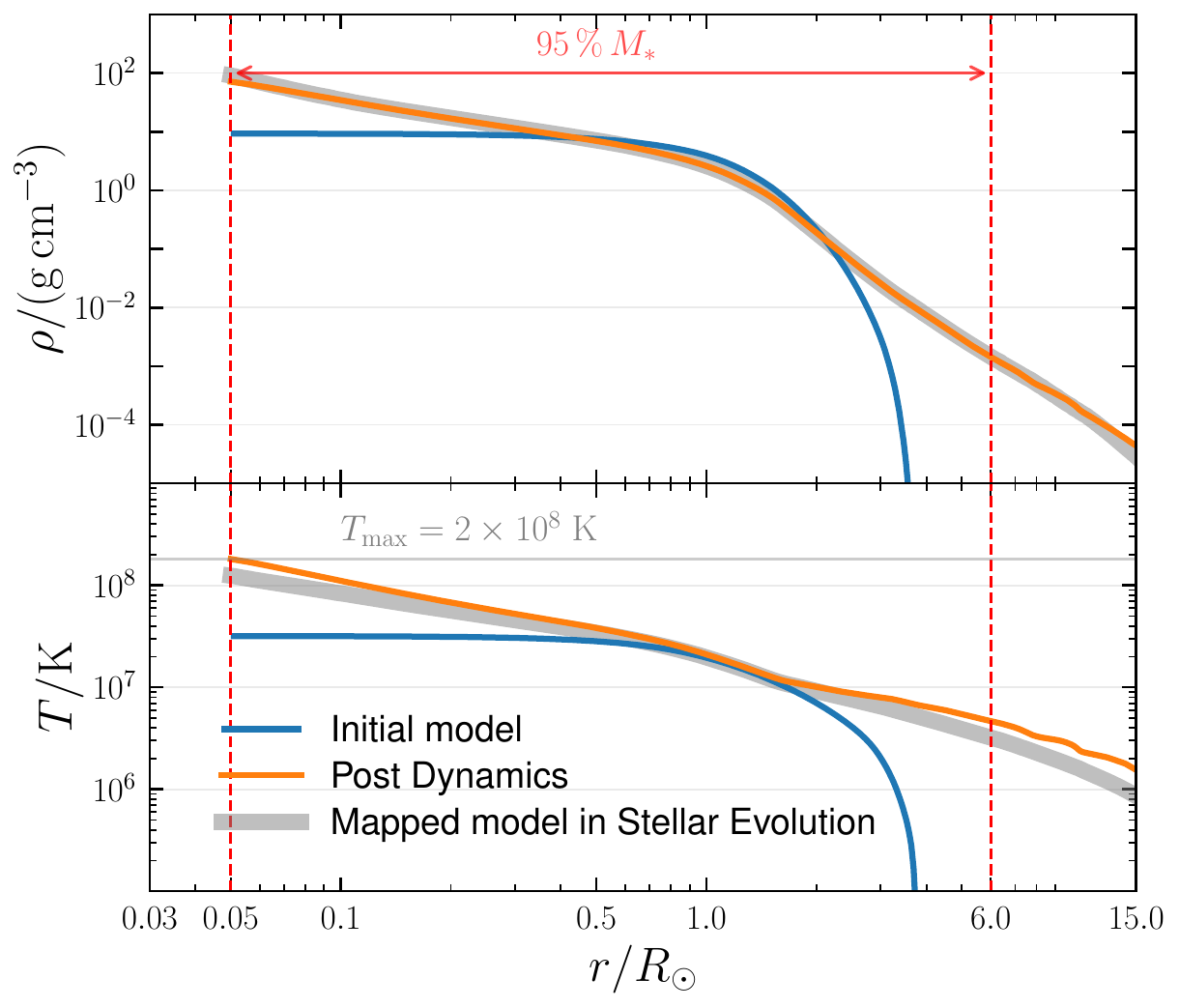}
    \caption{Comparison of radial density (upper panel) and temperature (lower panel) profiles using the 3D-to-1D mapping procedure.
    The blue and orange curves correspond to the initial and final profiles of the hydrodynamical simulation.
    The gray curves represent the mapped profiles (i.e., the initial model for the MESA calculation after the dynamical phase). The red dashed lines mark 
    the inner boundary of the stellar profile and 
    the radius enclosing $95\%$ of the stellar mass above the compact core.
    The horizontal gray line in the lower panel marks $2\times10^8\,$K.
}
    \label{fig:stellar profile}
\end{figure}

Figure \ref{fig:final-configuration} presents the top view (panel a) and side view (panel b) of the system at the end of the hydrodynamical simulation, and the profiles of the angle-averaged radial velocity (panel c) and angular velocity (panel d). 
The inner region of the system becomes nearly spherical, whereas the outer envelope remains highly asymmetric (top view of Fig.~\ref{fig:final-configuration}); in particular, a tail formed by the passage of the NS on the left side. 
In order to measure how close the system is to the hydrostatic equilibrium, we present the ratio of the angle-averaged radial velocity ($v_{\rm r}$) over Keplerian velocity at the equator ($v_{\rm Kep}$) in the panel (c) of Fig.~\ref{fig:final-configuration}, where $v_{\rm r}$ is given by
\be
v_{r}(r) = \frac{1}{4\pi}\int_{\sigma(r)} (\bm{v}\cdot \hat{\mathbf{r}}) \d\Omega. 
\ee
The ratio of $v_{\rm r}/v_{\rm Kep}$ remains below $3\%$ of the local Keplerian velocity throughout the structure, suggesting no significant radial motions, even in the outer region. Meanwhile, the envelope retains a certain amount of angular momentum transferred from the orbital motion of the NS during the hydrodynamical inspiral phase. In the panel (d) of Fig.~\ref{fig:final-configuration}, we present the ratio of the angular velocity ($\Omega_\ast$) over the Keplerian angular velocity at the equator ($\Omega_{\rm Kep}$), where $\Omega_{\ast}$ is obtained from the profile of  specific angular momentum (Eq.\,\ref{eq:avg-j}), which is   
\be
\Omega_\ast(r)= \frac{3}{2}\frac{ j_{\ast}(r)}{r^2},
\ee
where the factor $3/2$ comes from the moment of inertia of a rotating shell. The inner region of the system, which contains most of the mass ($95\% M_{\ast}$ within $6R_{\odot}$), is slowly rotating. While $\Omega_{\ast}/\Omega_{\rm Kep}$ reaches about 0.4 at $r=6\,R_\odot$, it is insufficient to form any disk-like structures (panel b of Fig.\,\ref{fig:final-configuration}). Therefore, the structure of the system is very close to the hydrostatic equilibrium and not considerably affected by the effects of rotation.

The 1D profiles obtained from this structure are shown in Fig.\,\ref{fig:stellar profile}.
Compared with the initial profile, the post-dynamics profile has a much denser and hotter core and an extended envelope.
The difference is mainly due to the gravity of the NS, which compresses the surrounding gas, forming a density spike at the inner region. 
At the same time, the compression significantly heats up the material near the NS. The temperature near the compact core ($2\times10^8\,$K; Fig.\,\ref{fig:stellar profile}) can be understood by comparing the internal energy to the  gravitational energy (i.e., virialized gas), 
which gives
\be
\begin{split}
  T_{\rm core} & \approx \xi\frac{GM_{\rm core}m_{\rm p}}{R_{\rm core}}k_{\rm B}^{-1}\\
&\approx 3.3 \times 10^8 {\rm K} \left(\frac{M_{\rm core}}{1.4M_{\odot}}\right)\left(\frac{R_{\rm core}}{0.05R_{\odot}}\right)^{-1}\left( \frac{\xi}{0.5}\right),  
\end{split}
\label{eq: core temp}
\ee
where $T_{\rm core}$ is the temperature at the compact core surface, $\xi$ is a coefficient related to the e.o.s of the envelope gas, $k_{\rm B}$ is the Boltzmann constant, and $m_{\rm p}$ is the proton mass.
This estimate roughly matches the innermost temperature shown in Fig.~\ref{fig:stellar profile}. However, we note that the core temperature depends on $R_{\rm core}$ and can be higher in the smaller region (cf.,  Sect.\ref{sec: discuss}).
The temperature is high near the compact core and remains at $\sim6\times10^6\mathrm{K}$ at the radius enclosing $95\%$ of the stellar mass (Fig.~\ref{fig:stellar profile}). Consequently, intense nuclear burning extends from the compact core surface through most of the envelope by mass, significantly affecting the subsequent evolution in MESA (see Sect.~\ref{sec: 1D stellar}).

In addition, we also compare the angle-averaged 1D profiles at the end of the hydrodynamical simulation and the profiles mapped into the MESA code (i.e., the starting point of the long-term evolution) in Fig.\,\ref{fig:stellar profile}.
While the density profiles well match each other, there are small differences in the temperature profiles.
The reason is that in the hydrodynamical simulation, the temperature is calculated by the ideal gas e.o.s in post-processing, while the mapped one by the realistic e.o.s in the MESA code. We do not expect this to have considerable effects on our results. 

\subsection{Stellar evolution}
\label{sec: 1D stellar}

\begin{figure*}[!tp]
    \centering
    \includegraphics[
    width=0.95\linewidth,
    height=0.65\textheight,
    keepaspectratio
    ]{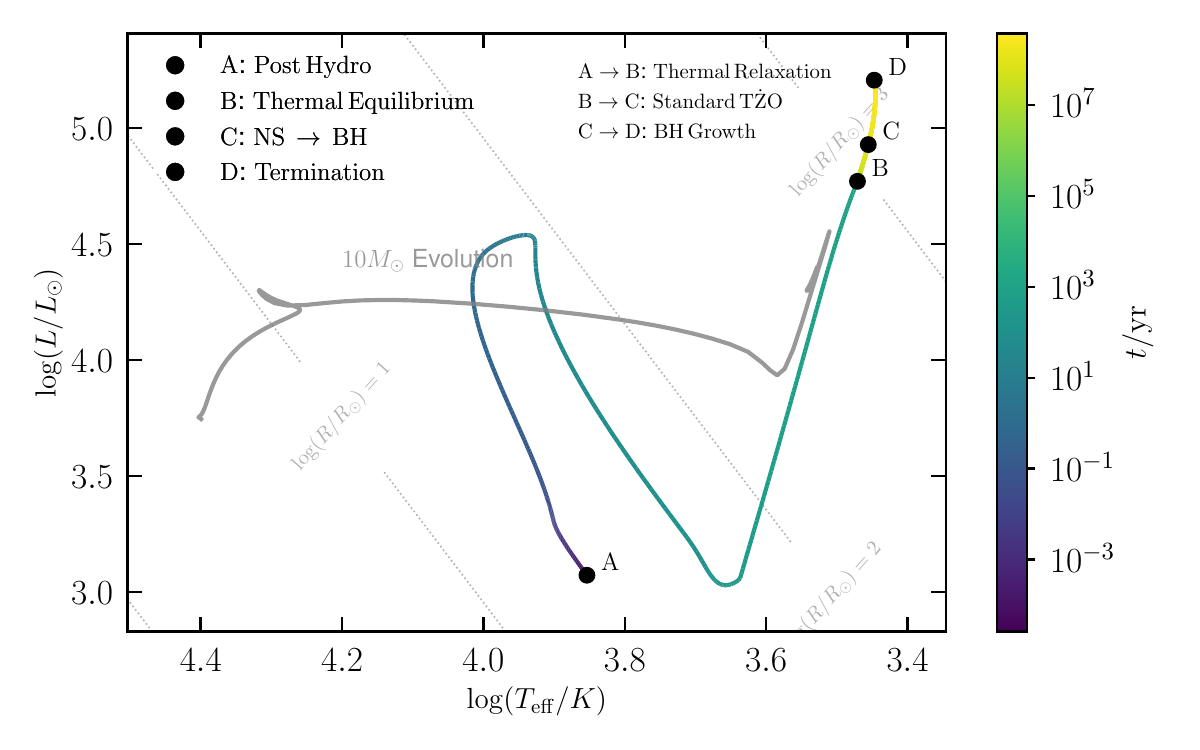}
    \caption{Hertzsprung–Russell (HR) diagram of the \tzo\ evolution. The colored track shows the long-term evolution of the \tzo, with color indicating the evolutionary time.  For comparison, the gray curve shows the evolutionary track of a non-rotating  single star of $10M_{\odot}$ from ZAMS to core helium depletion (with no stellar wind). Four representative evolutionary stages, labeled by A–D and marked by black filled circles, are highlighted along the \tzo\ track (see text for details). The diagonal gray dotted lines denote constant-radius lines corresponding to $\log(R/R_{\odot})$=1, 2, and 3. 
    }
    \label{fig:hr-diagram}
\end{figure*}

\begin{figure}[!tp]
    \centering
    \includegraphics[width=0.99\linewidth]{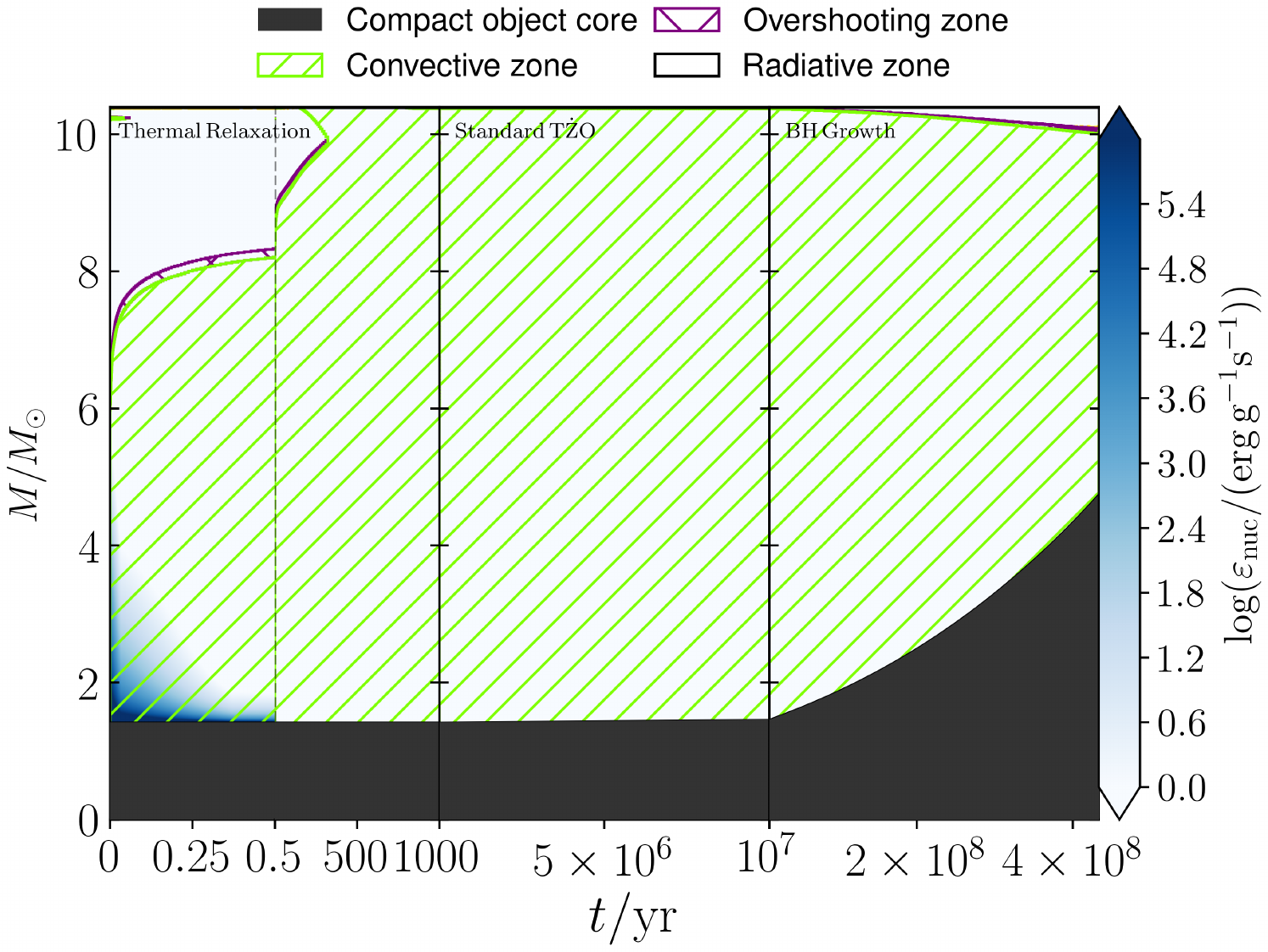}    \caption{Kippenhahn diagram of the long-term evolution of the reference \tzo\ model. The dark gray region represents the compact-object core. The hatched regions indicate convective (green) and overshooting (purple) zones, while unhatched regions within the envelope are radiative. The color map shows the local nuclear energy generation rate, expressed as $\log(\varepsilon_{\rm nuc}/(\mathrm{erg}\,\mathrm{g}^{-1}\,\mathrm{s}^{-1}))$. The evolutionary sequence is divided into the thermal-relaxation, standard \tzo, and BH-growth phases. During thermal relaxation, the dominant energy source shifts from nuclear burning to accretion.
    }
    \label{fig:kipp-diagram}
\end{figure}

\begin{figure}[!tp]
    \centering
    \includegraphics[width=0.99\linewidth]{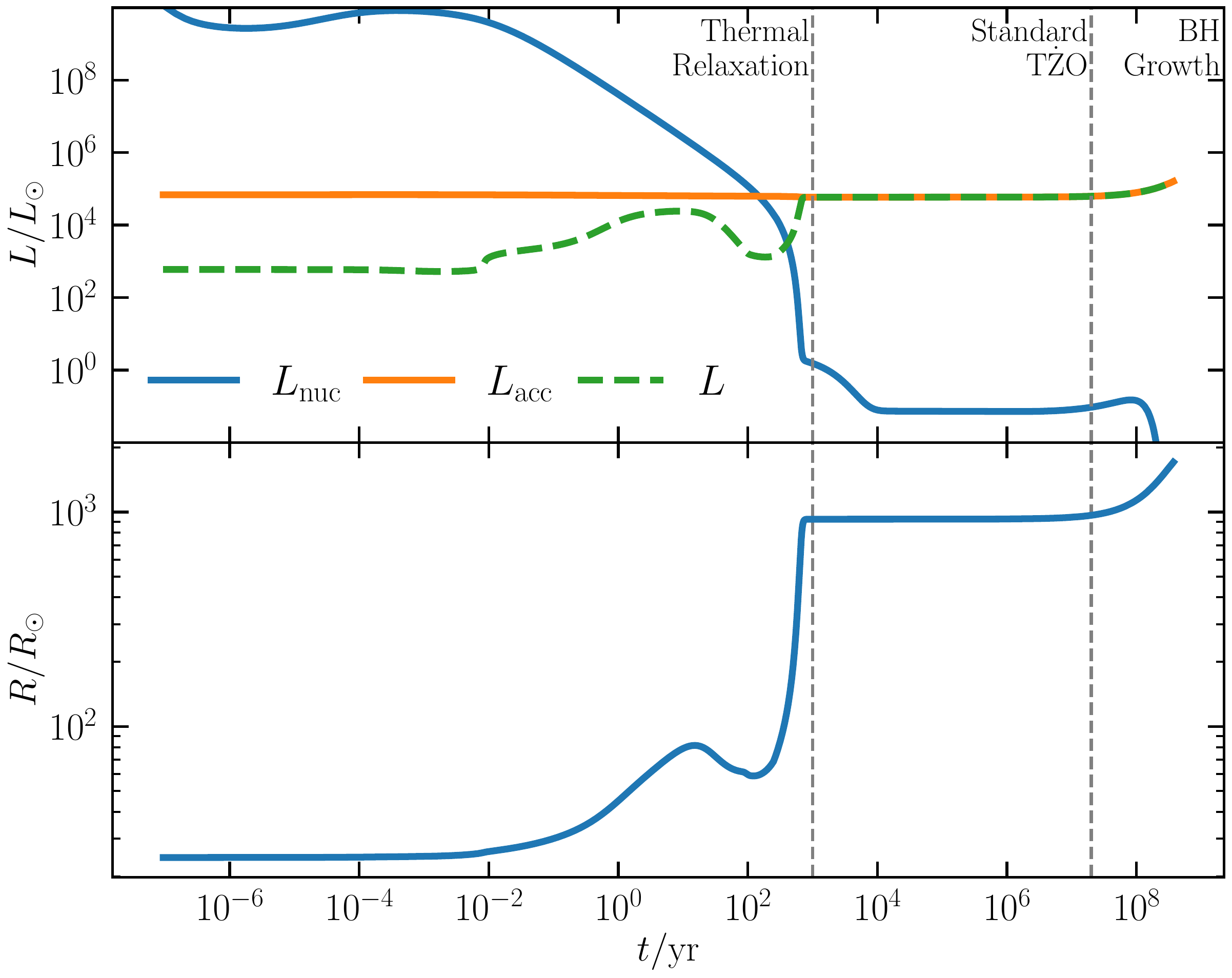}
    \caption{Thermal evolution of the reference \tzo\ model. The upper panel shows the nuclear luminosity $L_{\rm nuc}$ (blue), accretion luminosity $L_{\rm acc}$ (orange), and photosphere luminosity $L$ (green), and the lower panel shows the stellar radius $R/R_{\odot}$. The vertical dashed lines indicate different evolutionary phases, including thermal relaxation, standard \tzo\ and BH growth phases.}
    \label{fig:evolution}
\end{figure}

Figure\,\ref{fig:hr-diagram} presents the long-term evolution of our \tzo\ on the Hertzsprung–Russell (HR) diagram; we also show the track of a non-rotating single star model without stellar wind for comparison. The corresponding stellar structure is presented in Fig.~\ref{fig:kipp-diagram}, and the evolution of the radius and surface luminosity in Fig.~\ref{fig:evolution}. The evolution exhibits three major stages having different evolutionary timescales:
\begin{itemize}
    \item Thermal relaxation phase: after the dynamical assembly, the resulting \tzo\ is far from  the thermal equilibrium. In particular, the energy generation rate of nuclear reaction is significantly higher than the luminosity at the surface  $L$. The thermal relaxation phase lasts until the \tzo\ reaches the global thermal equilibrium,  i.e. when $L=L_{\rm acc}+ L_{\rm nuc}$.
    \item Standard T\.ZO phase:  Once the system reaches the thermal equilibrium, the system remains stationary for about $10^7\,$yr, and the compact core experiences negligible growth. Since the structure resembles the static solution explored in previous works (e.g., \citealp{1975ApJ...199L..19T, 1977ApJ...212..832T, 1991ApJ...380..167B}), we refer to this stage as the ``Standard T\.ZO" phase.
    \item Black-hole growth phase: the formation of a BH core is expected when the core mass reaches $2\,M_\odot$. After that, the compact core grows at an accelerated rate, which we call the ``BH growth" phase. 
\end{itemize}
In the following subsections, we describe the evolution during these stages in detail. 

\subsubsection{Thermal relaxation phase}

The entire thermal relaxation takes about 1000\,yr (A to B in Fig.\,\ref{fig:hr-diagram} and 0--1000\,yr in Figs.\,\ref{fig:kipp-diagram}-\ref{fig:evolution}). 
In the early stage, the outer stellar envelope is radiative, and the  luminosity and surface temperature increase as the star evolves (Fig.\,\ref{fig:hr-diagram}). This is because the nuclear power is initially about 6 orders of magnitude higher than the surface luminosity (Fig.\,\ref{fig:evolution}), driving the expansion of the entire star. As seen in Fig.~\ref{fig:kipp-diagram}, the nuclear burning efficiency then drops accordingly, making the burning region shrink toward the surface of the compact object core. Meanwhile, the cooling makes the envelope more opaque, leading to a shrinking radiative zone.
Once the star cools down to a certain extent, the nuclear power can no longer support the entire structure, which leads to a slight contraction of the star. Consequently, the accretion power of the NS core dominates the evolution of stellar structure making the surface luminosity adjust to the accretion power ($t=100\,$yr in Fig.\,\ref{fig:evolution}).
The star restarts the expansion at the same time, with the envelope becoming fully convective.  
Thermal equilibrium (i.e., $L_{\rm acc} \approx L$) is achieved at about 1000\,yr. This is consistent with the thermal timescale ($\tau_{\rm th}$), given by
\be
\begin{split}
\tau_{\rm th} &\sim \frac{U}{L} \approx \xi\frac{G(M_{\ast}+M_{\rm NS})^2}{R_{\ast}}\frac{1}{ L_{\rm acc}},    \\
& \approx 2 \times 10^3 \textrm{yr} \left( \frac{\xi}{0.5}\right)\left( \frac{M_{\ast}+M_{\rm NS}}{10.4 M_{\odot}}\right)^2 \left(\frac{R_{\ast}}{ 15 R_{\odot}}\right)^{-1} \\&\times\left(\frac{L_{\mathrm{acc}}}{5\times 10^4 L_{\odot}}\right)^{-1},
\end{split}
\ee
where $U$
is the total internal energy of the star estimated using the virial theorem.

\begin{figure}[!tp]
    \centering
    \includegraphics[width=0.95\linewidth]{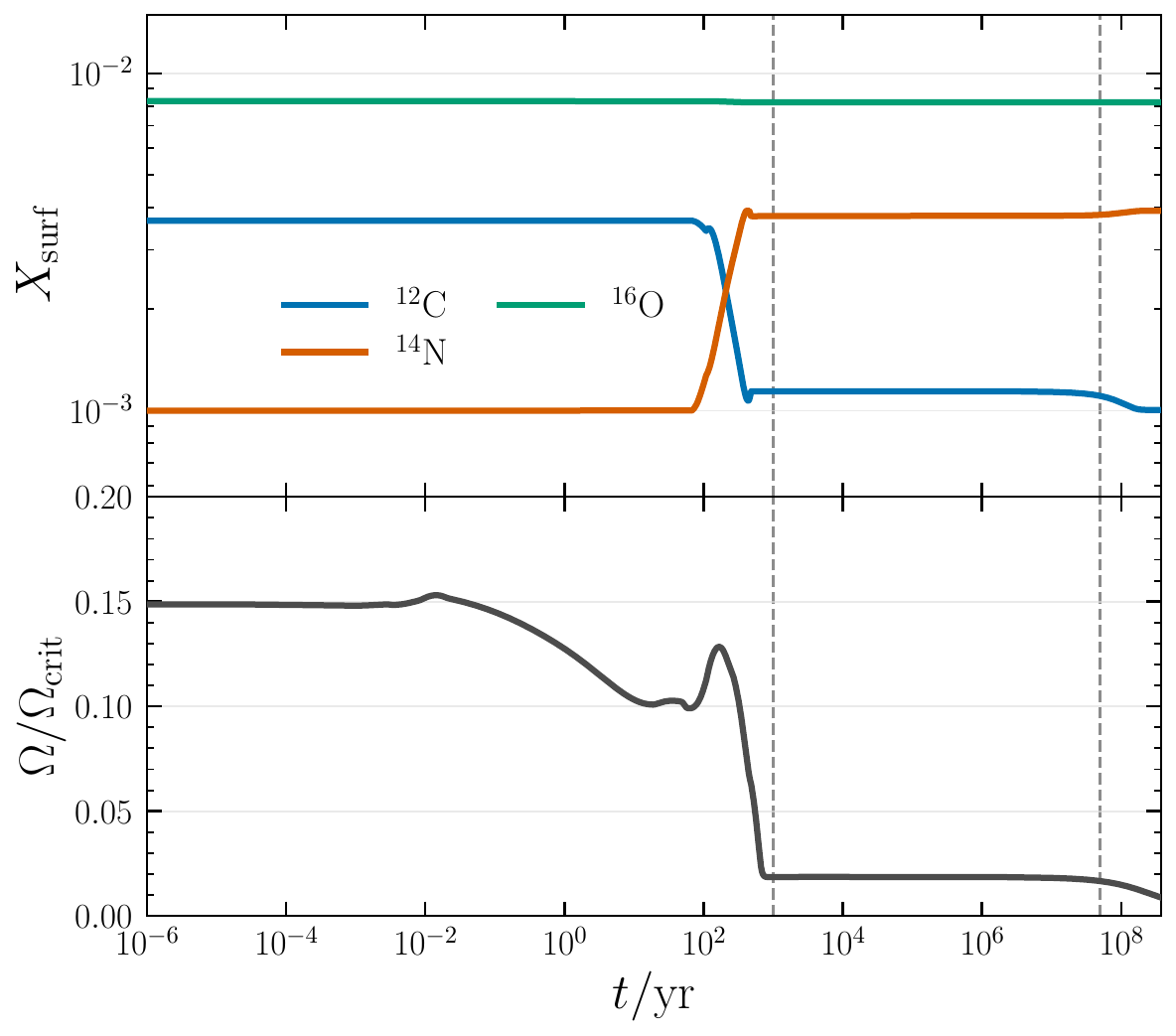}
    \caption{Chemical and rotational evolution of the reference \tzo\ model. The upper panel shows the surface mass fractions of $^{12}\mathrm{C}$, $^{14}\mathrm{N}$, and $^{16}\mathrm{O}$, while the lower panel shows the surface angular velocity normalized by its critical value, $\Omega/\Omega_{\rm crit}$, where $\Omega_{\rm crit}=[GM_{\rm tot}(1-\Gamma_{\rm Edd})/r^{3}]^{1/2}$ is the critical angular velocity, and $\Gamma_{\rm Edd}$ is the surface-averaged Eddington factor. The vertical dashed lines separate the thermal-relaxation, standard \tzo, and BH-growth phases.}
    \label{fig:accompany-evo}
\end{figure}

As shown in Fig.~\ref{fig:kipp-diagram}, intense nuclear burning initially occurs in a large fraction of the mass, and convection transports the CNO-cycle processed material to the surface, producing carbon depletion and nitrogen enhancement with little change in $^{16}\mathrm{O}$ (Fig.~\ref{fig:accompany-evo}). Since the temperature in these nuclear burning regions is significantly higher than that in a normal MS star (Fig.\ref{fig:stellar profile}), the corresponding abundance pattern is different from the canonical CNO cycle \footnote{For example, at a density of $100\,\mathrm{g\,cm^{-3}}$ and a temperature of $10^8\mathrm{K}$, the mass fraction of nitrogen is about 4 times that of carbon at equilibrium \citep{1972NPhS..238...23C}, which is about 140 times for the canonical CNO cycle. }. Once the envelope becomes fully convective, the processed material is well mixed with the unburnt envelope, resulting in the final surface abundance. These spatially extended, formation-induced nuclear burning regions differ from standard \tzo\ models, which may provide a complementary chemical signature for identifying \tzos\ through observations (Sect.~\ref{sec: discuss}).

In addition, we also present the evolution of the surface angular velocity (in units of the critical angular velocity) in Fig.~\ref{fig:accompany-evo}. The expansion during the thermal relaxation stage substantially reduces its rotational rate. By the time thermal equilibrium is reached, the rotation rate has fallen below $5\%$ of the critical value and has little effect on the subsequent evolution (see App.~\ref{app:rotation}). We therefore ignore the effect of rotation in the following discussion.

\subsubsection{Standard T\.ZO phase}

This phase corresponds to the track from B to C on the HR diagram shown in Fig.~\ref{fig:hr-diagram}.  
Since our reference model does not take into account stellar wind, the evolution after reaching the thermal equilibrium is driven by the growth of the compact core. The growth rate is determined by the Eddington accretion rate (Eq.\,\ref{eq:Mdot-edd}), and it takes about $10^7\,$yr for the NS core to reach $2M_\odot$ and become a BH. 
Consistent with previous studies \citep[e.g.,][]{1992ApJ...386..206C, 1993MNRAS.263..817C}, the envelope supported by accretion power is fully convective, with a stellar radius of about $10^3\,R_{\odot}$. The surface luminosity is $6.3\times 10^4L_{\odot}$, matching the scaling law given in Cannon (\citeyear{1993MNRAS.263..817C}). 

\subsubsection{Black-hole growth phase}

After the NS becomes a BH, the mass of the BH grows on a timescale of about $10^{8}\,$yr.
The evolutionary track on the HR diagram remains near the Hayashi line due to its strongly convective envelope (C to D in Fig.\,\ref{fig:hr-diagram}). Different from a normal post-MS star, the structure of our model is supported by the accretion power, allowing it to be brighter than the maximal luminosity expected by the single star model of the same mass. 
With the growth of the BH core, the envelope mass is reduced, and the Eddington factor in the envelope increases.
At the end of the simulation ($t\approx 4\times 10^8\,$yr), 
the radiation pressure dominates over the gas pressure with the Eddington factor largely exceeding 1, which the MESA code cannot handle. We discuss the implications of this configuration in Sect.\,\ref{sec: final fate}. 

\subsection{Model uncertainties}
\label{sec: para}

\begin{figure*}[!tp]
    \centering
    \includegraphics[
        width=0.90\linewidth,
        height=0.65\textheight,
        keepaspectratio
    ]{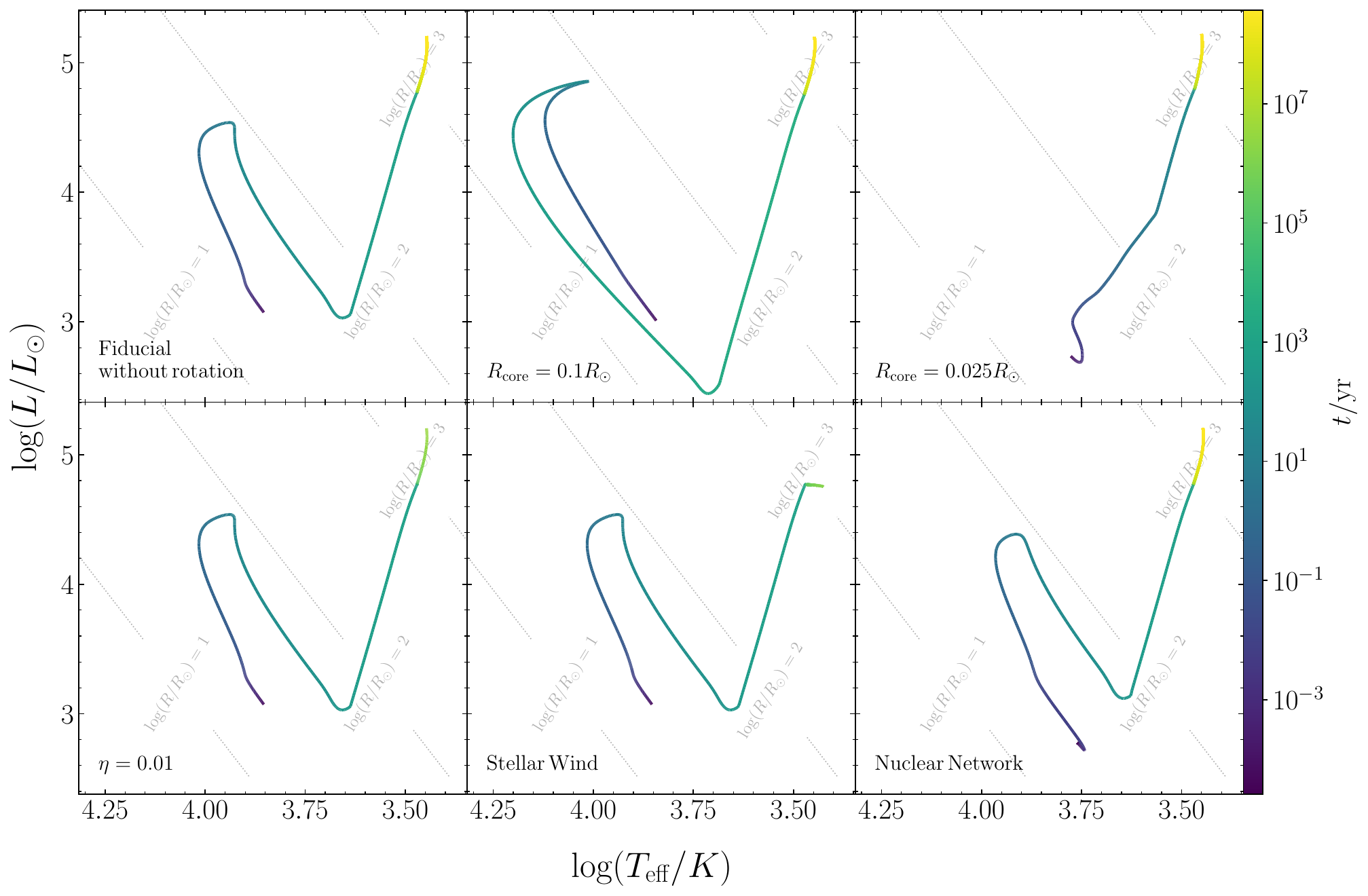}
    \caption{Hertzsprung--Russell (HR) diagrams comparing the long-term evolution of \tzos\ under different choices of physical assumptions. The top panels show models with different inner core radii, $R_{\rm core}=0.05\,R_{\odot}$ (reference), $0.1\,R_{\odot}$, and $0.025\,R_{\odot}$, respectively. The bottom panels show the effects of a reduced accretion efficiency ($\eta=0.01$), the inclusion of stellar winds, and an alternative nuclear reaction network. Grey dotted lines denote location of constant stellar radius. The color-coded tracks trace the long-term evolution of each model, with color indicating evolutionary time.}
    \label{fig:hr-dif}
\end{figure*}

In this subsection, we investigate how the uncertainties in the input physics affect our result. 
We vary the size of the compact core ($R_{\rm core}$), the mass accretion rate of the compact core (controlled by $\eta$), wind mass loss, and nuclear network. 
In the following computations, we turn off stellar rotation for simplicity, since the evolution without rotation on the HR diagram (fiducial model without rotation; see the upper left panel in Fig.\,\ref{fig:hr-dif}) does not show considerable difference with the one with rotation (fiducial model; see Fig.~\ref{fig:hr-diagram}). All the models are computed until the break of hydrostatic equilibrium. 

\subsubsection{Size of the compact core}

Our calculation cannot resolve the structure around the NS (radius about 10 km $\sim 10^{-5} R_{\odot}$). Here we address this uncertainty by varying the core radius $R_{\rm core}$ from $0.025\,R_\odot$ to $0.1\,R_\odot$. As shown in Fig.~\ref{fig:hr-dif}, the choice of $R_{\rm core}$ only affects the evolution during thermal relaxation, while the evolution after reaching thermal equilibrium remains unchanged. 
The value of $R_{\rm core}$ determines the temperature around the compact object core  (Eq.~\ref{eq: core temp}), affecting the nuclear power of the star. In the $R_{\rm core}=0.025R_\odot$ case, the nuclear power is strong enough to drive a continuous expansion, while the other cases show a brief contraction due to the rapid decay of the nuclear power. Stronger nuclear burning may therefore process more envelope material
and produce greater surface enrichment. We emphasize that enrichment
is a robust qualitative outcome of our models, without attempting
to predict its magnitude quantitatively.

\subsubsection{Accretion rate}

In the reference model, we compute the mass accretion rate by taking $\eta=1$ in Eq.\,\eqref{eq:Mdot-edd}. 
However, the accretion rate can be much higher if the accretion flow is cooled by neutrino emission \citep{1993ApJ...411L..33C,2012ApJ...752L...2C}.
To address this uncertainty, we vary $\eta$ from 1 down to 0.1 and 0.01 (only the $\eta=0.01$ case is presented in Fig.\,\ref{fig:hr-dif}). The evolutionary tracks appear to be identical to the reference case. The difference lies in the evolutionary timescale, as the compact core grows at different rates with different $\eta$. For $\eta=1$, the final age of the model is $\sim100$\,Myr, but $\sim1$\,Myr for $\eta=0.01$.

\subsubsection{Stellar wind}

\begin{figure}[!tp]
    \centering
    \includegraphics[trim=0.5cm 0.5cm 0.5cm 0.5cm,clip,width=0.99\linewidth]{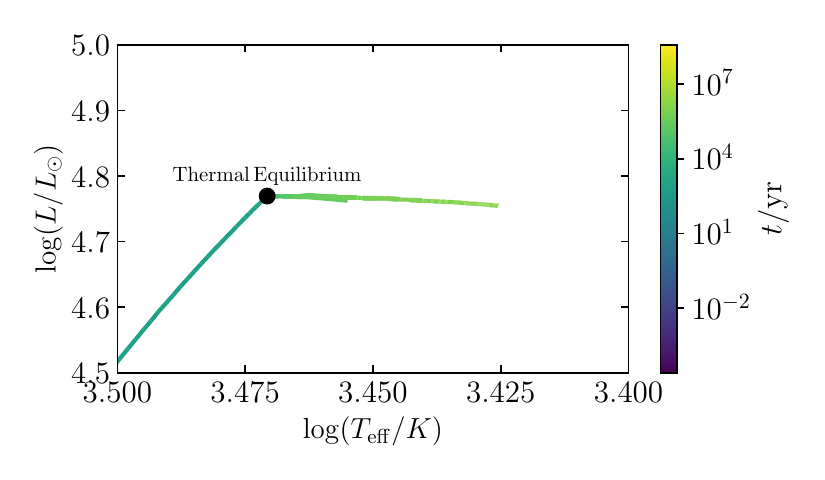}
    \caption{Evolutionary track of the model in the HR diagram during the wind–loss phase. The black dot represents the model entering thermal equilibrium, the same as point B in our fiducial model.
    \label{fig:hrd-wind-zoomin}
    }
\end{figure}

In the fiducial model, we do not consider the effects of stellar wind for simplicity, which may describe the long-term evolution of \tzos\ at low metallicity. At high metallicity, strong stellar winds can significantly alter the evolution of massive stars. We account for stellar winds by recomputing the reference model (no rotation) with the wind prescription determined by \citet{Brott2011A&A...530A.115B}.
The thermal relaxation phase persists, since the timescale is too short for a significant wind mass loss. In the following long-term evolution, the stellar envelope is stripped by winds within about 3\,Myr, leaving a naked neutron star, and the mass of the compact core remains roughly unchanged during the evolution. In Fig.\,\ref{fig:hrd-wind-zoomin}, we provide a zoom-in for the evolution after reaching the thermal equilibrium. The luminosity remains roughly constant, since the structure is supported by the accretion power of the compact core. 

\subsubsection{Nuclear Network}

Our fiducial model adopts the simplified nuclear network \texttt{approx21.net}, which is designed to capture the main hydrogen-burning processes using approximate reaction pathways, while omitting many intermediate reactions and species. It does not provide a complete treatment of high-temperature burning processes, such as the Ne--Na and Mg--Al cycles. As shown in Eq.~\eqref{eq: core temp} and Fig.~\ref{fig:stellar profile}, the inner region reaches temperatures well above those required for standard hydrogen burning, allowing these additional reactions to occur. Although this hot-burning region contains only a small fraction of the stellar mass and has limited impact on the overall stellar structure, it may still leave behind a chemical signature through the evolution process.

We recompute the reference (no rotation) model with a larger network, \texttt{stern.net} tested in \citet{2024A&A...690A.135J}, which contains 77 reactions, including not only the CNO cycle but also the Ne-Na and Mg-Al
cycles, and the burnings of helium, carbon, neon, and oxygen. The initial metallicity remains unchanged but with the metal distribution by \citet{Asplund2021A&A...653A.141A} to initialize extra elements.
We see that the evolutionary track computed using stern.net is similar to the fiducial model, but the resulting chemical abundances are slightly different. 

\section{Discussion}
\label{sec: discuss}

\subsection{Caveats of our assumptions}
\label{sec: assump}

We start our hydrodynamical simulation with a non-rotating $10M_{\odot}$ ZAMS star. However, after the SN explosion, the companion would first interact with the SN ejecta. The impact can strip a small amount of envelope material and deposit thermal energy into the star, causing the envelope to inflate \citep{Liu:2015tya,Hirai:2025oyi, Cary2026ApJ...996..141C}, enhancing the probability of the NS-MS collision \citep[e.g.,][]{Ercolino2026arXiv260521062E}. However, this process is sensitive to many factors, like explosion energy and ejecta mass, which is beyond the scope of this work. Importantly, the impact of the SN ejecta mainly affects the outer envelope, leaving the inner stellar structure nearly unchanged \citep{2021MNRAS.505.2485O}. Since the inflated envelope only contains a small fraction of the stellar mass, we do not expect this uncertainty to considerably affect our results.

In the classical \tzo\ models, energy transport in the innermost region around the NS core is dominated by radiative diffusion instead of convection. The transition region between the radiative and convective zones is known as the ``knee", located at a radius of $\sim100\,$km \citep{1977ApJ...212..832T}. This knee feature plays a crucial role in determining the structure and energy source of \tzos, and in massive \tzos, the interrupted rapid-proton reaction occurring around the knee produces unique chemical signatures \citep{1991ApJ...380..167B,1993MNRAS.263..817C}. While our MESA simulation cannot resolve this knee region due to the resolution limit of the hydrodynamical simulation, the properties of the standard \tzo\ stage are broadly consistent with previous studies (cf., Sect.\,\ref{sec: 1D stellar}), indicating that our treatment is adequate for investigating the structure of the envelope.  

\subsection{Comparison with classical T\.ZO Evolution}
\label{sec: comparsion}

By construction, the energy generation of classical T\.ZO models is dominated by the innermost region around the NS core \citep{1977ApJ...212..832T,1991ApJ...380..167B,1993MNRAS.263..817C}. However, we demonstrate that, for the T\.ZOs emerging from NS-MS collisions,  a large fraction of the mass is initially hot enough to initiate hydrogen burning (Fig.\,\ref{fig:kipp-diagram}), and the corresponding nuclear power can be orders of magnitude higher than the Eddington luminosity of the NS core (Fig.\,\ref{fig:evolution}). This condition leads to a stage of thermal relaxation, which is not explored in previous studies, and the evolution during this stage is sensitive to the choice of $R_{\rm core}$ (cf., Sect.\,\ref{sec: para}). In particular, the contribution of nuclear burning increases with decreasing $R_{\rm core}$ (Fig.\,\ref{fig: para-evo}). While the structure of our T\.ZO model is supported by the accretion power after the thermal relaxation stage, it is unclear whether the structure becomes nuclear-power-supported if our calculation can resolve deeper regions. 

In classical \tzo\ models, there is no stable solution for the T\.ZO mass in the range 8--14$\,M_\odot$ \citep[e.g.,][]{1993MNRAS.263..817C}. This is because, in this mass range, pair production around the knee region becomes important, altering the opacity and therefore reducing the Eddington luminosity (Eq.\,\ref{eq: Ledd}), and the nuclear power near the NS core is not strong enough to compensate this decrease in the accretion luminosity. However, we show that the core region is not the only significant energy source and that nuclear burning in the overlying envelope also contributes to supporting the envelope. Consequently, the thermal relaxation process may adjust the nuclear power in the envelope to compensate the decrease in the accretion luminosity, filling the mass gap expected by the classical \tzo\ models.

\subsection{Final fate of \tzos}
\label{sec: final fate}

As described in Sect.\,\ref{meth: 1D stellar}, our long-term evolution runs until the crash of the MESA calculation, indicating the breakdown of the hydrostatic equilibrium. At the end of the calculation, the Eddington ratio in the stellar envelope typically exceeds unity. For example, in the reference model, the Eddington factor is over 50 at a mass coordinate near $9\,M_\odot$ (Fig.\,\ref{fig: eddington}). Such a configuration can also occur in massive MS stars but with a lower maximal Eddington factor \citep{Grafener2012A&A...538A..40G,Sanyal2015A&A...580A..20S}.
In our model, the radiation energy stored beneath the outer envelope
can exceed the gravitational binding energy of the overlying material,
providing a potential energy source for eruptive mass ejections
\citep[e.g.,][]{Cheng2024ApJ...974..270C}.
Radiative heat exchange can excite pulsations and convert part of
this energy into the kinetic energy of the gas.
Our stability analysis identifies a growing radial mode, suggesting
that these pulsations may become strong enough to unbind and eject
the outer layers (Appendix~\ref{app:eddington}). Therefore, we expect that, before becoming an isolated NS or BH, a T\.ZO may experience mass ejections similar to luminous blue variables near the end of its life. 

Our calculation does not include any possible energy release associated with the NS-to-BH transition. This is reasonable when most material and energy are trapped by the newly formed event horizon \citep[e.g.,][]{Baiotti2005PhRvD..71b4035B}. However, if the progenitor NS has a large amount of rotational angular momentum, a disk can form during the collapse of the NS, producing jets \citep{Margalit2015PhRvL.115q1101M}. It is unclear whether such jets can break out from the envelope and produce transient events.  

In addition, if nuclear reaction is the dominant energy source, T\.ZOs may end up with a violent explosion \citep{1995MNRAS.274..485P, 2018MNRAS.475L..49M, 2021MNRAS.508...74M}. When such T\.ZOs exhaust their nuclear fuel, gravitational collapse is expected, as the accretion power is not strong enough to support their envelopes. 

\subsection{Observation signatures}
\label{sec: obs}

T\.ZOs can be observationally identified by their unusual surface abundance.
Previous searches focus on lithium enrichment and the products of 
the interrupted proton-capture process \citep{2014MNRAS.445L..36T,2018MNRAS.479.3101B,2021A&A...646A..98T}. However, lithium enrichment through the Cameron-Fowler process is not unique for T\.ZOs \citep[e.g.,][]{2018MNRAS.479.3101B,2021A&A...646A..98T}. 
In addition to these signatures, our results imply that the elements related to the CNO cycle (even Ne-Na, and Mg-Al cycles) may also help with identifying T\.ZOs. During the NS-MS collision, stellar material is significantly heated by shocks. Thus, intense nuclear reaction occurs in a large fraction of the mass (cf., Fig.\,\ref{fig:kipp-diagram}), causing considerable nitrogen and $[^{13}\mathrm{C}/^{12}\mathrm{C}]$ enrichment at the surface.   

While the interrupted proton-capture process is expected to occur in massive T\.ZOs, the enrichment in nitrogen and $[^{13}\mathrm{C}/^{12}\mathrm{C}]$ can occur in less massive T\.ZOs as long as the stellar material is significantly heated during the formation process. However, these signatures are not unique to T\.ZOs either. For example, rotational mixing can also transport CNO-cycle processed material to the surface \citep{Brott2011A&A...530A.115B}, and the hot-bottom burning in super-asymptotic giant branch (sAGB) stars can produce carbon depletion and nitrogen enhancement at the surface \citep{Doherty2014MNRAS.437..195D}. 
In addition, the chemical signatures emerging from our models may also occur in a higher-mass regime, potentially appearing as chemically peculiar red supergiants.

\section{Conclusion}
\label{sec: conc}

In this work, we have investigated the formation of a \tzo\ through NS-MS collision and its long-term evolution. We focus on a 10 $M_\odot$ MS star. By combining three-dimensional hydrodynamical simulations with one-dimensional stellar-evolution calculations, we connect the dynamical formation of a \tzo\ to its subsequent thermal and nuclear evolution.   
The whole evolution can be summarized as follows:

\begin{itemize}

\item
The 3D hydrodynamical simulations show that an NS colliding with an MS star can lead to the formation of a bound remnant that rapidly settles into a quasi-spherical configuration. In our simulation, roughly $90\%$ of the original stellar mass remains bound within $R_\ast\approx15\,R_\odot$, with the NS located at the density maximum. This dynamically formed remnant provides the initial structure for the subsequent stellar-evolution calculations.

\item
The newly formed \tzo\ is initially far from thermal equilibrium and undergoes an early thermal-relaxation phase lasting approximately $10^3\,\mathrm{yr}$. Compression and shocks during the collision phase heat up a substantial fraction of the star, producing an initial nuclear luminosity much larger than the surface luminosity. The envelope consequently expands and cools, becomes fully convective, and eventually approaches thermal equilibrium with the surface luminosity $L$ equal to the accretion luminosity $L_{\rm acc}$. The products of the nuclear reaction are transported to the surface by convection, producing carbon depletion and nitrogen enhancement.

\item
After thermal relaxation, the system enters a long-lived ``standard \tzo'' phase. With weak winds and accretion at the Eddington rate, the object has a cool, fully convective envelope with $R_\ast\approx10^3\,R_\odot$ and $L\simeq6.3\times10^4\,L_\odot$, and remains near the Hayashi line for $\gtrsim10^7\,\mathrm{yr}$. The global structure during this stage broadly resembles the equilibrium \tzo\ solutions explored in previous studies.

\item
At later times, continued growth of the compact core and the transition from an NS to a BH make the envelope increasingly radiation-pressure dominated. Our fiducial model terminates when part of the envelope becomes strongly super-Eddington, suggesting the onset of substantial or eruptive mass loss. The present calculations do not directly determine the subsequent envelope disruption. When stellar winds are included, the envelope is instead stripped within approximately $3\,\mathrm{Myr}$, leaving a nearly naked NS before substantial core growth, and the envelope also becomes super-Eddington in the end. 

\item
Our exploration with various model parameters shows that the early thermal-relaxation phase and its associated nucleosynthesis are sensitive to the unresolved compact-core boundary. By contrast, the global evolution after thermal equilibrium is comparatively insensitive to the inner-boundary radii and nuclear-reaction networks. The accretion efficiency primarily changes the evolutionary timescale, whereas stellar-wind mass loss can qualitatively alter the lifetime and final fate of the system.

\end{itemize}

The key distinction between our model and classical \tzo\ models concerns the origin of the surface chemical anomalies. In standard models, these anomalies are produced by nuclear burning confined to the innermost region surrounding the NS core. In our dynamically formed \tzo, the NS--MS collision heats up a much larger fraction of the star through shocks and compression, leading to widespread nuclear burning during the thermal-relaxation phase. Convection subsequently transports the processed material to the surface, producing carbon depletion and nitrogen enhancement. 

After thermal relaxation, the stellar structure approaches a classical
accretion-supported \tzo\ configuration, while the surface composition
can retain a chemical imprint of the earlier nuclear burning.
The resulting carbon and nitrogen anomalies may help identify \tzo\
candidates when combined with other observational properties, although
similar abundance patterns are likely to occur in other stars. More detailed nuclear networks and better-resolved calculations of the compact-core region will be required to obtain quantitative abundance predictions.

\begin{acknowledgments}

We thank Philipp Podsiadlowski and Liang Dai for insightful discussions. 
We also thank Charlie Gibson and Qinghui Sun for helpful technical advice and assistance. 
The simulations presented in this work were performed on the TDLI-Astro cluster at the Tsung-Dao Lee Institute. 
X.T.X. was supported by the National Natural Science Foundation of China (grant No. 12603055).

\end{acknowledgments}

\appendix

\section{Further details of the Reference model}

\subsection{Rotation}
\label{app:rotation}

In this appendix, we examine the rotational evolution of the \tzo\ and its impact on stellar evolution. Tidal interactions before the SN may reduce the MS star's spin toward synchronization; for simplicity, we adopt an initially non-rotating MS star, so the newly formed \tzo\ acquires its angular momentum primarily from the post-SN orbit. During thermal relaxation, expansion slows the rotation, interrupted by a brief contraction-induced spin-up, while internal transport redistributes angular momentum within the envelope. By the time thermal equilibrium is reached, the surface angular velocity falls below $0.05\Omega_{\rm crit}$ (Fig.~\ref{fig:accompany-evo}).

Following \mesa, we define
$\Omega_{\rm crit}=\Omega_{\rm Kep}(1-\Gamma_{\rm Edd})^{1/2}$,
where $\Omega_{\rm Kep}=(GM_{\rm tot}/R_{\ast}^{3})^{1/2}$ and
$M_{\rm tot}=M_{\rm core}+M_{\ast}$ (Sect.~\ref{sec: core}).
Here, $\Gamma_{\rm Edd}=\kappa L/(4\pi cGM_{\rm tot})$ is the surface Eddington factor, evaluated using the surface-layer opacity and the total stellar mass, rather than the compact-core mass used in Eq.~\eqref{eq: Ledd}. We use this definition during the hydrostatic, sub-Eddington phase.

The rotation rate after thermal relaxation can be estimated from the initial orbital angular momentum,
$J_{\rm orb}=\mu[G(M_{\ast}^{\rm ini}+M_{\rm NS})a(1-e^2)]^{1/2}$,
where $\mu=M_{\ast}^{\rm ini}M_{\rm NS}/(M_{\ast}^{\rm ini}+M_{\rm NS})$,
and $a$ and $e$ are the post-SN orbital parameters.
Approximating the \tzo\ as a rigidly rotating body gives
$I\Omega=f_JJ_{\rm orb}$, where $f_J$ is the retained fraction of orbital angular momentum and $I=k_I M_{\rm tot}R_{\ast}^{2}$, with $k_I$ the dimensionless moment-of-inertia coefficient. Thus,
\be
\label{eq:rot-general}
\frac{\Omega}{\Omega_{\rm Kep}} = \frac{f_J}{k_I}\frac{\mu}{M_{\rm tot}}\left(\frac{M_{\ast}^{\rm ini}+M_{\rm NS}}{M_{\rm tot}}\right)^{1/2}\left[\frac{a(1-e^{2})}{R_{\ast}}\right]^{1/2}.
\ee

For the reference system, $M_{\ast}^{\rm ini}=10\,M_{\odot}$ and $M_{\rm NS}=1.4\,M_{\odot}$, while the bound envelope has $M_{\ast}\simeq9\,M_{\odot}$ after the collision, giving $M_{\rm tot}\simeq10.4\,M_{\odot}$. Equation~\eqref{eq:rot-general} becomes
\be
\label{eq:rot}
\frac{\Omega}{\Omega_{\rm Kep}} \simeq 0.018\,f_J\left(\frac{k_I}{2/5}\right)^{-1}\left(\frac{a}{9\,R_{\odot}}\right)^{1/2}\left(\frac{1-e^{2}}{0.36}\right)^{1/2}\left(\frac{R_{\ast}}{10^{3}\,R_{\odot}}\right)^{-1/2}.
\ee
Taking $f_J=1$ and $k_I=2/5$ (a uniform-density sphere), the reference values $a=9\,R_{\odot}$, $e=0.8$, and $R_{\ast}=10^3\,R_{\odot}$ give $\Omega/\Omega_{\rm Kep}\simeq0.018$.
The normalization used in Fig.~\ref{fig:accompany-evo} is instead
$\Omega/\Omega_{\rm crit}=(\Omega/\Omega_{\rm Kep})(1-\Gamma_{\rm Edd})^{-1/2}$.
This estimate illustrates the reduction in rotation as the star expands, but neglects the actual density profile, angular-momentum loss, and differential rotation.

For the same illustrative parameters, the centrifugal acceleration relative to gravity at the stellar surface is
\be
\label{eq:rot-acceleration}
\frac{a_{\rm cent}}{g} = \frac{\Omega^2R_{\ast}}{GM_{\rm tot}/R_{\ast}^{2}} = \left(\frac{\Omega}{\Omega_{\rm Kep}}\right)^2 \simeq3\times10^{-4}.
\ee
The small influence of rotation is also supported by the non-rotating calculation, whose HR-diagram track is nearly indistinguishable from that of the rotating reference model (Sect.~\ref{sec: para}; Fig.~\ref{fig:hr-dif}). Rotation therefore has little effect on the subsequent evolution in our models or the principal conclusions of this work.

\subsection{The final profile of the Eddington factor}
\label{app:eddington}
\begin{figure*}[!tp]
    \centering
    \includegraphics[
        width=0.7\linewidth,
        height=0.65\textheight,
        keepaspectratio
    ]{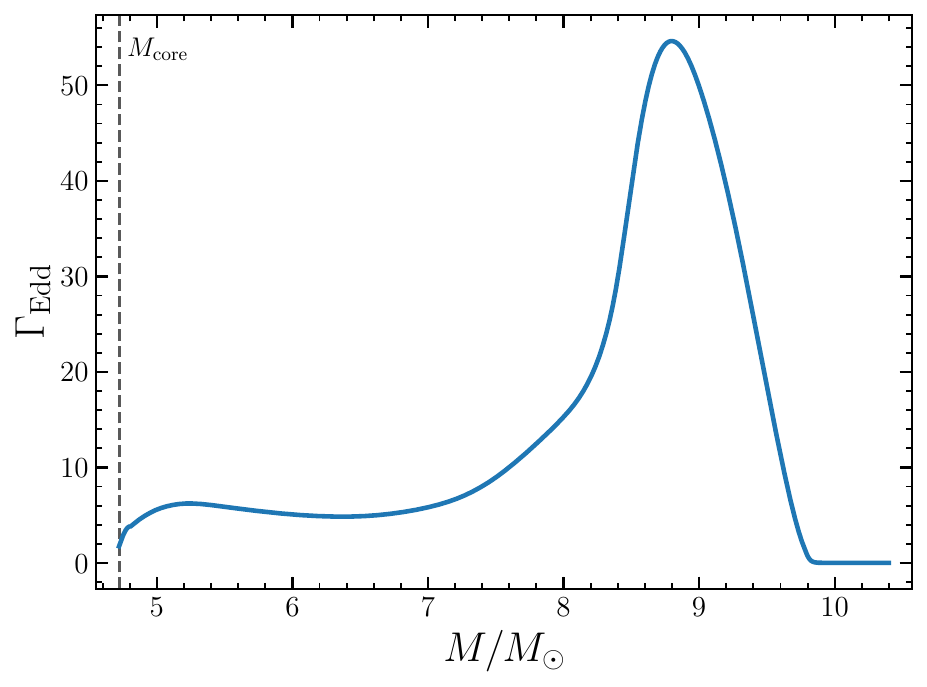}
    \caption{Final profile of the local Eddington factor, $\Gamma_{\rm Edd}$, as a function of enclosed mass for the last converged model in our reference sequence. The vertical dashed line marks the BH-core boundary at $M_{\rm core}$.}
    \label{fig: eddington}
\end{figure*}

Figure~\ref{fig: eddington} shows the local Eddington-factor profile of the last converged reference model during the BH-growth phase. Here, $\Gamma_{\rm Edd}=L/L_{\rm Edd}$ uses the total local luminosity, with $L_{\rm Edd}$ evaluated from the local opacity and enclosed mass. The vertical dashed line marks the BH-core boundary at $M_{\rm core}$. Throughout most of the envelope, $\Gamma_{\rm Edd}>1$, reaching a maximum of $\simeq55$ at an enclosed mass of $\simeq8.8\,M_\odot$, before decreasing sharply in the outermost layers.

To investigate why our MESA calculation terminates, we first
compare the radiation energy beneath the outer envelope with
the gravitational binding energy of the overlying material.
We divide the final profile at $m_b\simeq8.80\,M_\odot$,
the peak of the total-luminosity ratio $L/L_{\rm Edd}$,
and define the radiation energy of the inner envelope as
$E_{\rm rad,in}=\int_{M_{\rm BH}}^{m_b}(3P_{\rm rad}/\rho)\,dm$
and the gravitational binding energy of the outer envelope as
$E_{\rm grav,out}=\int_{m_b}^{M_{\rm tot}}(Gm/r)\,dm$.
The ratio $E_{\rm rad,in}/E_{\rm grav,out}\simeq1.34$ shows
that the stored radiation provides an energy reservoir large
enough in principle to unbind these outer layers, provided
sufficient energy is transferred to them.

To further assess the stability of our final model,
we perform a linear perturbation analysis.
\textsc{GYRE} is a numerical tool for stellar pulsation and stability analysis,
integrated with MESA
\citep{TownsendTeitler2013,paxton:15}.
We use version 8.1
to examine the stability of non-adiabatic
radial perturbations throughout the envelope of our final model,
assuming that perturbations to heat transport are dominated by radiation.
\textsc{GYRE} retains radial acceleration in the momentum equation,
\begin{equation}
    \frac{\partial^2 r}{\partial t^2}
    =-4\pi r^2\frac{\partial P}{\partial m}-\frac{Gm}{r^2},
    \label{eq:radial_momentum}
\end{equation}
where $P$ includes gas and radiation pressure.
We adopt a time dependence $e^{-i\omega t/t_{\rm dyn}}$, where
$\omega$ is the dimensionless
complex frequency and $t_{\rm dyn}=[R_\ast^3/(GM_{\rm tot})]^{1/2}$.
\textsc{GYRE} solves the linearized stellar evolution equations
for $\omega$ in our final MESA model.

We find a growing radial mode with
$\omega\simeq0.741+0.115i$, indicating that the final profile
is pulsationally unstable on a timescale of approximately
$10\,{\rm yr}$.
Together with the energy budget, this supports our interpretation
that the model is approaching mass ejection, which would prevent
further hydrostatic evolution.
\clearpage
\section{Models with different \texorpdfstring{$R_{\mathrm{\lowercase{core}}}$}{Rcore}}

\begin{figure*}[!htbp]
    \centering
    \includegraphics[
        width=0.8\linewidth,
        height=0.65\textheight,
        keepaspectratio
    ]{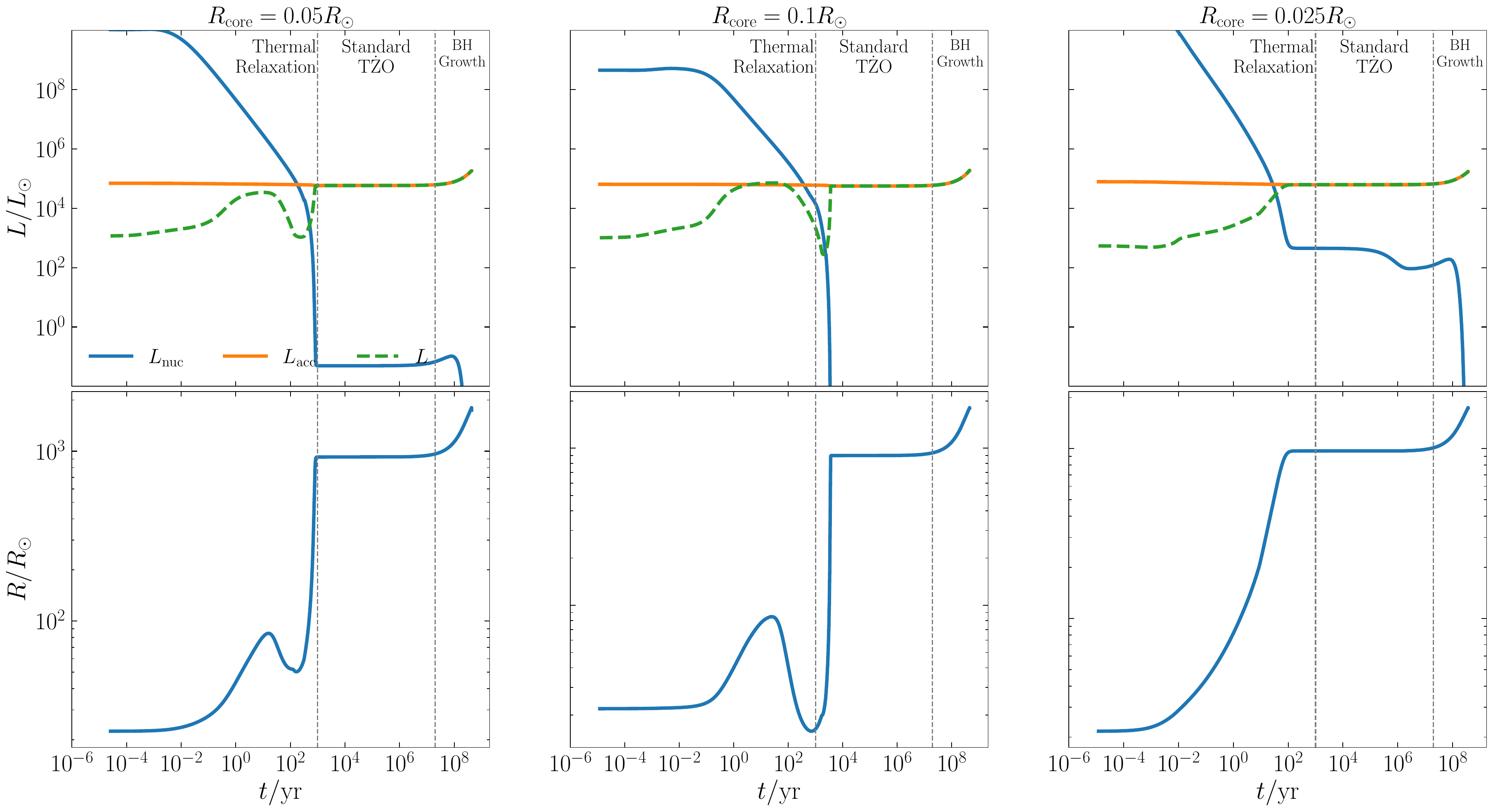}
    \caption{Dependence of the long-term T$\dot{\rm Z}$O evolution on the
    adopted compact-core radius. From left to right, the columns show models
    with $R_{\rm core}=0.05$, $0.1$, and $0.025\,R_{\odot}$, respectively;
    the left column corresponds to the reference model. The upper panels show
    the nuclear luminosity $L_{\rm nuc}$ (blue), accretion luminosity
    $L_{\rm acc}$ (orange), and total surface luminosity $L$ (green dashed).
    The lower panels show the evolution of the stellar radius $R_{\ast}$.
    The vertical dotted lines delimit the thermal-relaxation, standard
    T$\dot{\rm Z}$O, and black-hole-growth phases. The compact-core radius
    mainly affects the evolution during thermal relaxation, whereas the
    post-relaxation surface luminosity and stellar radius remain broadly
    similar among the three models.}
    \label{fig: para-evo}
\end{figure*}

\clearpage

\bibliographystyle{aasjournalv7}
\bibliography{reference}

\begin{thebibliography}{}
\expandafter\ifx\csname natexlab\endcsname\relax\def\natexlab#1{#1}\fi
\providecommand{\url}[1]{\href{#1}{#1}}
\providecommand{\dodoi}[1]{doi:~\href{http://doi.org/#1}{\nolinkurl{#1}}}
\providecommand{\doeprint}[1]{\href{http://ascl.net/#1}{\nolinkurl{http://ascl.net/#1}}}
\providecommand{\doarXiv}[1]{\href{https://arxiv.org/abs/#1}{\nolinkurl{https://arxiv.org/abs/#1}}}

\bibitem[{P. Anninos {et~al.}(2026)Anninos, Portman, Carmichael, Hoffman, \&
  Sieverding}]{Anninos:2025fti}
Anninos, P., Portman, M.~E., Carmichael, S.~R., Hoffman, R.~D., \& Sieverding,
  A. 2026, \bibinfo{title}{{r-process Nucleosynthesis from Hyperaccreting
  Neutron Stars in Common Envelopes},} Astrophys. J., 997, 88,
  \dodoi{10.3847/1538-4357/ae2311}

\bibitem[{M. {Asplund} {et~al.}(2021){Asplund}, {Amarsi}, \&
  {Grevesse}}]{Asplund2021A&A...653A.141A}
{Asplund}, M., {Amarsi}, A.~M., \& {Grevesse}, N. 2021, \bibinfo{title}{{The
  chemical make-up of the Sun: A 2020 vision},} \aap, 653, A141,
  \dodoi{10.1051/0004-6361/202140445}

\bibitem[{L. {Baiotti} {et~al.}(2005){Baiotti}, {Hawke}, {Montero},
  {L{\"o}ffler}, {Rezzolla}, {Stergioulas}, {Font}, \&
  {Seidel}}]{Baiotti2005PhRvD..71b4035B}
{Baiotti}, L., {Hawke}, I., {Montero}, P.~J., {et~al.} 2005,
  \bibinfo{title}{{Three-dimensional relativistic simulations of rotating
  neutron-star collapse to a Kerr black hole},} \prd, 71, 024035,
  \dodoi{10.1103/PhysRevD.71.024035}

\bibitem[{E.~R. {Beasor} {et~al.}(2018){Beasor}, {Davies}, {Cabrera-Ziri}, \&
  {Hurst}}]{2018MNRAS.479.3101B}
{Beasor}, E.~R., {Davies}, B., {Cabrera-Ziri}, I., \& {Hurst}, G. 2018,
  \bibinfo{title}{{A critical re-evaluation of the Thorne-{\.Z}ytkow object
  candidate HV 2112},} \mnras, 479, 3101, \dodoi{10.1093/mnras/sty1744}

\bibitem[{G.~T. {Biehle}(1991){Biehle}}]{1991ApJ...380..167B}
{Biehle}, G.~T. 1991, \bibinfo{title}{{High-Mass Stars with Degenerate Neutron
  Cores},} \apj, 380, 167, \dodoi{10.1086/170572}

\bibitem[{E.
  {B{\"o}hm-Vitense}(1958){B{\"o}hm-Vitense}}]{Bohm-Vitense1958ZA.....46..108B}
{B{\"o}hm-Vitense}, E. 1958, \bibinfo{title}{{{\"U}ber die
  Wasserstoffkonvektionszone in Sternen verschiedener Effektivtemperaturen und
  Leuchtkr{\"a}fte. Mit 5 Textabbildungen},} \zap, 46, 108

\bibitem[{I. {Brott} {et~al.}(2011){Brott}, {de Mink}, {Cantiello}, {Langer},
  {de Koter}, {Evans}, {Hunter}, {Trundle}, \&
  {Vink}}]{Brott2011A&A...530A.115B}
{Brott}, I., {de Mink}, S.~E., {Cantiello}, M., {et~al.} 2011,
  \bibinfo{title}{{Rotating massive main-sequence stars. I. Grids of
  evolutionary models and isochrones},} \aap, 530, A115,
  \dodoi{10.1051/0004-6361/201016113}

\bibitem[{C. Brummel-Smith {et~al.}(2019)Brummel-Smith, Bryan, Butsky, Corlies,
  Emerick, Forbes, Fujimoto, Goldbaum, Grete, Hummels, Kim, Koh, Li, Li, Li,
  OShea, Peeples, Regan, Salem, Schmidt, Simpson, Smith, Tumlinson, Turk, Wise,
  Abel, Bordner, Cen, Collins, Crosby, Edelmann, Hahn, Harkness, Harper-Clark,
  Kong, Kritsuk, Kuhlen, Larrue, Lee, Meece, Norman, Oishi, Paschos, Peruta,
  Razoumov, Reynolds, Silvia, Skillman, Skory, So, Tasker, Wagner, Wang, Xu, \&
  Zhao}]{Brummel-Smith2019}
Brummel-Smith, C., Bryan, G., Butsky, I., {et~al.} 2019, \bibinfo{title}{ENZO:
  An Adaptive Mesh Refinement Code for Astrophysics (Version 2.6),} Journal of
  Open Source Software, 4, 1636, \dodoi{10.21105/joss.01636}

\bibitem[{R.~C. {Cannon}(1993){Cannon}}]{1993MNRAS.263..817C}
{Cannon}, R.~C. 1993, \bibinfo{title}{{Massive Thorne-{\.Z}ytkow Objects -
  Structure and Nucleosynthesis},} \mnras, 263, 817,
  \dodoi{10.1093/mnras/263.4.817}

\bibitem[{R.~C. {Cannon} {et~al.}(1992){Cannon}, {Eggleton}, {Zytkow}, \&
  {Podsiadlowski}}]{1992ApJ...386..206C}
{Cannon}, R.~C., {Eggleton}, P.~P., {Zytkow}, A.~N., \& {Podsiadlowski}, P.
  1992, \bibinfo{title}{{The Structure and Evolution of Thorne-Zytkow
  Objects},} \apj, 386, 206, \dodoi{10.1086/171006}

\bibitem[{S. {Cary} {et~al.}(2026){Cary}, {Lu}, {Leung}, \&
  {Wong}}]{Cary2026ApJ...996..141C}
{Cary}, S., {Lu}, W., {Leung}, C., \& {Wong}, T. L.~S. 2026,
  \bibinfo{title}{{Accretion from a Shock-inflated Companion: Spinning Down
  Neutron Stars to Hour-long Periods},} \apj, 996, 141,
  \dodoi{10.3847/1538-4357/ae1d43}

\bibitem[{G.~R. Caughlan \& W.~A. Fowler(1972)Caughlan \&
  Fowler}]{1972NPhS..238...23C}
Caughlan, G.~R., \& Fowler, W.~A. 1972, \bibinfo{title}{Mean Lifetimes and
  Equilibrium Abundances in the Fast {CN} Cycle,} Nature Physical Science, 238,
  23, \dodoi{10.1038/physci238023a0}

\bibitem[{S.~J. {Cheng} {et~al.}(2024){Cheng}, {Goldberg}, {Cantiello},
  {Bauer}, {Renzo}, \& {Conroy}}]{Cheng2024ApJ...974..270C}
{Cheng}, S.~J., {Goldberg}, J.~A., {Cantiello}, M., {et~al.} 2024,
  \bibinfo{title}{{A Model for Eruptive Mass Loss in Massive Stars},} \apj,
  974, 270, \dodoi{10.3847/1538-4357/ad701e}

\bibitem[{P.~C.-K. {Cheong} {et~al.}(2026{\natexlab{a}}){Cheong}, {Fryer}, \&
  {Radice}}]{2026arXiv260809395C}
{Cheong}, P. C.-K., {Fryer}, C.~L., \& {Radice}, D. 2026{\natexlab{a}},
  \bibinfo{title}{{Hyperaccreting Magnetised Neutron Stars inside Rotating
  Massive Envelopes: Low-Power Jets and Precursor Flares},} arXiv e-prints,
  arXiv:2608.09395, \dodoi{10.48550/arXiv.2608.09395}

\bibitem[{P.~C.-K. {Cheong} {et~al.}(2026{\natexlab{b}}){Cheong}, {Radice}, \&
  {Fryer}}]{2026arXiv260423503C}
{Cheong}, P. C.-K., {Radice}, D., \& {Fryer}, C.~L. 2026{\natexlab{b}},
  \bibinfo{title}{{Hyperaccreting Neutron Stars inside Massive Envelopes: The
  Implausibility of Thorne-{\.Z}ytkow Objects},} \prd, 114, 023018,
  \dodoi{10.1103/8pb2-9t2x}

\bibitem[{R. Chevalier(1995)Chevalier}]{CHEVALIER199595}
Chevalier, R. 1995, \bibinfo{title}{Neutron star accretion in dense
  environments,} Physics Reports, 256, 95, \dodoi{10.1016/0370-1573(94)00103-A}

\bibitem[{R.~A. {Chevalier}(1993){Chevalier}}]{1993ApJ...411L..33C}
{Chevalier}, R.~A. 1993, \bibinfo{title}{{Neutron Star Accretion in a Stellar
  Envelope},} \apjl, 411, L33, \dodoi{10.1086/186905}

\bibitem[{R.~A. {Chevalier}(1996){Chevalier}}]{1996ApJ...459..322C}
{Chevalier}, R.~A. 1996, \bibinfo{title}{{Neutrino-cooled Accretion: Rotation
  and Stellar Equation of State},} \apj, 459, 322, \dodoi{10.1086/176895}

\bibitem[{R.~A. {Chevalier}(2012){Chevalier}}]{2012ApJ...752L...2C}
{Chevalier}, R.~A. 2012, \bibinfo{title}{{Common Envelope Evolution Leading to
  Supernovae with Dense Interaction},} \apjl, 752, L2,
  \dodoi{10.1088/2041-8205/752/1/L2}

\bibitem[{C.~L. {Doherty} {et~al.}(2014){Doherty}, {Gil-Pons}, {Lau},
  {Lattanzio}, \& {Siess}}]{Doherty2014MNRAS.437..195D}
{Doherty}, C.~L., {Gil-Pons}, P., {Lau}, H. H.~B., {Lattanzio}, J.~C., \&
  {Siess}, L. 2014, \bibinfo{title}{{Super and massive AGB stars - II.
  Nucleosynthesis and yields - Z = 0.02, 0.008 and 0.004},} \mnras, 437, 195,
  \dodoi{10.1093/mnras/stt1877}

\bibitem[{A. {Ercolino} {et~al.}(2026){Ercolino}, {Langer}, {Gal-Yam},
  {Schootemeijer}, {Mannes}, {Jin}, {Valli}, {de Mink}, \&
  {Dessart}}]{Ercolino2026arXiv260521062E}
{Ercolino}, A., {Langer}, N., {Gal-Yam}, A., {et~al.} 2026,
  \bibinfo{title}{{Neutron star-companion interaction in core collapse
  supernovae. Population synthesis based on detailed binary evolution models},}
  arXiv e-prints, arXiv:2605.21062.
\newblock \doarXiv{2605.21062}

\bibitem[{R.~W. {Everson} {et~al.}(2024){Everson}, {Hutchinson-Smith},
  {Vigna-G{\'o}mez}, \& {Ramirez-Ruiz}}]{2024ApJ...971..132E}
{Everson}, R.~W., {Hutchinson-Smith}, T., {Vigna-G{\'o}mez}, A., \&
  {Ramirez-Ruiz}, E. 2024, \bibinfo{title}{{Rethinking Thorne─{\.Z}ytkow
  Object Formation: Assembly via Common Envelope in Field Binaries},} \apj,
  971, 132, \dodoi{10.3847/1538-4357/ad595e}

\bibitem[{R. {Farmer} {et~al.}(2023){Farmer}, {Renzo}, {G{\"o}tberg},
  {Bellinger}, {Justham}, \& {de Mink}}]{2023MNRAS.524.1692F}
{Farmer}, R., {Renzo}, M., {G{\"o}tberg}, Y., {et~al.} 2023,
  \bibinfo{title}{{Observational predictions for Thorne-{\.Z}ytkow objects},}
  \mnras, 524, 1692, \dodoi{10.1093/mnras/stad1977}

\bibitem[{C.~L. {Fryer} {et~al.}(1996){Fryer}, {Benz}, \&
  {Herant}}]{1996ApJ...460..801F}
{Fryer}, C.~L., {Benz}, W., \& {Herant}, M. 1996, \bibinfo{title}{{The Dynamics
  and Outcomes of Rapid Infall onto Neutron Stars},} \apj, 460, 801,
  \dodoi{10.1086/177011}

\bibitem[{G. {Gr{\"a}fener} {et~al.}(2012){Gr{\"a}fener}, {Owocki}, \&
  {Vink}}]{Grafener2012A&A...538A..40G}
{Gr{\"a}fener}, G., {Owocki}, S.~P., \& {Vink}, J.~S. 2012,
  \bibinfo{title}{{Stellar envelope inflation near the Eddington limit.
  Implications for the radii of Wolf-Rayet stars and luminous blue variables},}
  \aap, 538, A40, \dodoi{10.1051/0004-6361/201117497}

\bibitem[{R. {Hirai} \& P. {Podsiadlowski}(2022){Hirai} \&
  {Podsiadlowski}}]{2022MNRAS.517.4544H}
{Hirai}, R., \& {Podsiadlowski}, P. 2022, \bibinfo{title}{{Neutron stars
  colliding with binary companions: formation of hypervelocity stars, pulsar
  planets, bumpy superluminous supernovae and Thorne-{\.Z}ytkow objects},}
  \mnras, 517, 4544, \dodoi{10.1093/mnras/stac3007}

\bibitem[{R. Hirai {et~al.}(2025)Hirai, Podsiadlowski, Hoeflich, Barkov, Chan,
  Liptai, \& Nagataki}]{Hirai:2025oyi}
Hirai, R., Podsiadlowski, P., Hoeflich, P., {et~al.} 2025,
  \bibinfo{title}{{Supernova-induced Binary-interaction-powered Supernovae: A
  Model for SN2022jli},} Astrophys. J., 995, 55,
  \dodoi{10.3847/1538-4357/ae172e}

\bibitem[{T. {Hutchinson-Smith} {et~al.}(2024){Hutchinson-Smith}, {Everson},
  {Twum}, {Batta}, {Yarza}, {Law-Smith}, {Vigna-G{\'o}mez}, \&
  {Ramirez-Ruiz}}]{2024ApJ...977..196H}
{Hutchinson-Smith}, T., {Everson}, R.~W., {Twum}, A.~A., {et~al.} 2024,
  \bibinfo{title}{{Rethinking Thorne─{\.Z}ytkow Object Formation: The Fate of
  X-Ray Binary LMC X-4 and Implications for Ultra-long Gamma-Ray Bursts},}
  \apj, 977, 196, \dodoi{10.3847/1538-4357/ad88f3}

\bibitem[{N. {Ivanova} {et~al.}(2013){Ivanova}, {Justham}, {Chen}, {De Marco},
  {Fryer}, {Gaburov}, {Ge}, {Glebbeek}, {Han}, {Li}, {Lu}, {Marsh},
  {Podsiadlowski}, {Potter}, {Soker}, {Taam}, {Tauris}, {van den Heuvel}, \&
  {Webbink}}]{Ivanova2013}
{Ivanova}, N., {Justham}, S., {Chen}, X., {et~al.} 2013,
  \bibinfo{title}{{Common envelope evolution: where we stand and how we can
  move forward},} \aapr, 21, 59, \dodoi{10.1007/s00159-013-0059-2}

\bibitem[{A.~S. {Jermyn} {et~al.}(2023){Jermyn}, {Bauer}, {Schwab}, {Farmer},
  {Ball}, {Bellinger}, {Dotter}, {Joyce}, {Marchant}, {Mombarg}, {Wolf}, {Sunny
  Wong}, {Cinquegrana}, {Farrell}, {Smolec}, {Thoul}, {Cantiello}, {Herwig},
  {Toloza}, {Bildsten}, {Townsend}, \& {Timmes}}]{jermyn22}
{Jermyn}, A.~S., {Bauer}, E.~B., {Schwab}, J., {et~al.} 2023,
  \bibinfo{title}{{Modules for Experiments in Stellar Astrophysics (MESA):
  Time-dependent Convection, Energy Conservation, Automatic Differentiation,
  and Infrastructure},} \apjs, 265, 15, \dodoi{10.3847/1538-4365/acae8d}

\bibitem[{H. {Jin} {et~al.}(2024){Jin}, {Langer}, {Lennon}, \&
  {Proffitt}}]{2024A&A...690A.135J}
{Jin}, H., {Langer}, N., {Lennon}, D.~J., \& {Proffitt}, C.~R. 2024,
  \bibinfo{title}{{Boron depletion in Galactic early B-type stars reveals two
  different main sequence star populations},} \aap, 690, A135,
  \dodoi{10.1051/0004-6361/202450896}

\bibitem[{P.~J.~T. {Leonard} {et~al.}(1994){Leonard}, {Hills}, \&
  {Dewey}}]{1994ApJ...423L..19L}
{Leonard}, P. J.~T., {Hills}, J.~G., \& {Dewey}, R.~J. 1994, \bibinfo{title}{{A
  New Way to Make Thorne-Zytkow Objects},} \apjl, 423, L19,
  \dodoi{10.1086/187225}

\bibitem[{Z.-W. Liu {et~al.}(2015)Liu, Tauris, R{\"o}pke, Moriya, Kruckow,
  Stancliffe, \& Izzard}]{Liu:2015tya}
Liu, Z.-W., Tauris, T.~M., R{\"o}pke, F.~K., {et~al.} 2015,
  \bibinfo{title}{{The interaction of core-collapse supernova ejecta with a
  companion star},} Astron. Astrophys., 584, A11,
  \dodoi{10.1051/0004-6361/201526757}

\bibitem[{M. {MacLeod} \& E. {Ramirez-Ruiz}(2015{\natexlab{a}}){MacLeod} \&
  {Ramirez-Ruiz}}]{2015ApJ...798L..19M}
{MacLeod}, M., \& {Ramirez-Ruiz}, E. 2015{\natexlab{a}}, \bibinfo{title}{{On
  the Accretion-fed Growth of Neutron Stars during Common Envelope},} \apjl,
  798, L19, \dodoi{10.1088/2041-8205/798/1/L19}

\bibitem[{M. {MacLeod} \& E. {Ramirez-Ruiz}(2015{\natexlab{b}}){MacLeod} \&
  {Ramirez-Ruiz}}]{2015ApJ...803...41M}
{MacLeod}, M., \& {Ramirez-Ruiz}, E. 2015{\natexlab{b}},
  \bibinfo{title}{{Asymmetric Accretion Flows within a Common Envelope},} \apj,
  803, 41, \dodoi{10.1088/0004-637X/803/1/41}

\bibitem[{B. {Margalit} {et~al.}(2015){Margalit}, {Metzger}, \&
  {Beloborodov}}]{Margalit2015PhRvL.115q1101M}
{Margalit}, B., {Metzger}, B.~D., \& {Beloborodov}, A.~M. 2015,
  \bibinfo{title}{{Does the Collapse of a Supramassive Neutron Star Leave a
  Debris Disk?},} \prl, 115, 171101, \dodoi{10.1103/PhysRevLett.115.171101}

\bibitem[{T.~J. {Moriya}(2018){Moriya}}]{2018MNRAS.475L..49M}
{Moriya}, T.~J. 2018, \bibinfo{title}{{Explosions of Thorne-{\.Z}ytkow
  objects},} \mnras, 475, L49, \dodoi{10.1093/mnrasl/sly005}

\bibitem[{T.~J. {Moriya} \& S.~I. {Blinnikov}(2021){Moriya} \&
  {Blinnikov}}]{2021MNRAS.508...74M}
{Moriya}, T.~J., \& {Blinnikov}, S.~I. 2021, \bibinfo{title}{{Properties of
  Thorne-{\.Z}ytkow object explosions},} \mnras, 508, 74,
  \dodoi{10.1093/mnras/stab2584}

\bibitem[{M. {Ogata} {et~al.}(2021){Ogata}, {Hirai}, \&
  {Hijikawa}}]{2021MNRAS.505.2485O}
{Ogata}, M., {Hirai}, R., \& {Hijikawa}, K. 2021,
  \bibinfo{title}{{Observability of inflated companion stars after supernovae
  in massive binaries},} \mnras, 505, 2485, \dodoi{10.1093/mnras/stab1439}

\bibitem[{A. {O'Grady} {et~al.}(2026){O'Grady}, {Moriya}, {Renzo}, \&
  {Vigna-G{\'o}mez}}]{2026enap....3..336O}
{O'Grady}, A., {Moriya}, T.~J., {Renzo}, M., \& {Vigna-G{\'o}mez}, A. 2026,
  \bibinfo{title}{{Thorne-{\.Z}ytkow objects},} in Encyclopedia of
  Astrophysics, Volume 3, Vol.~3, 336--357,
  \dodoi{10.1016/B978-0-443-21439-4.00093-6}

\bibitem[{B. {Paxton} {et~al.}(2011){Paxton}, {Bildsten}, {Dotter}, {Herwig},
  {Lesaffre}, \& {Timmes}}]{paxton:11}
{Paxton}, B., {Bildsten}, L., {Dotter}, A., {et~al.} 2011,
  \bibinfo{title}{{Modules for Experiments in Stellar Astrophysics (MESA)},}
  \apjs, 192, 3, \dodoi{10.1088/0067-0049/192/1/3}

\bibitem[{B. {Paxton} {et~al.}(2013){Paxton}, {Cantiello}, {Arras}, {Bildsten},
  {Brown}, {Dotter}, {Mankovich}, {Montgomery}, {Stello}, {Timmes}, \&
  {Townsend}}]{paxton:13}
{Paxton}, B., {Cantiello}, M., {Arras}, P., {et~al.} 2013,
  \bibinfo{title}{{Modules for Experiments in Stellar Astrophysics (MESA):
  Planets, Oscillations, Rotation, and Massive Stars},} \apjs, 208, 4,
  \dodoi{10.1088/0067-0049/208/1/4}

\bibitem[{B. {Paxton} {et~al.}(2015){Paxton}, {Marchant}, {Schwab}, {Bauer},
  {Bildsten}, {Cantiello}, {Dessart}, {Farmer}, {Hu}, {Langer}, {Townsend},
  {Townsley}, \& {Timmes}}]{paxton:15}
{Paxton}, B., {Marchant}, P., {Schwab}, J., {et~al.} 2015,
  \bibinfo{title}{{Modules for Experiments in Stellar Astrophysics (MESA):
  Binaries, Pulsations, and Explosions},} \apjs, 220, 15,
  \dodoi{10.1088/0067-0049/220/1/15}

\bibitem[{B. {Paxton} {et~al.}(2018){Paxton}, {Schwab}, {Bauer}, {Bildsten},
  {Blinnikov}, {Duffell}, {Farmer}, {Goldberg}, {Marchant}, {Sorokina},
  {Thoul}, {Townsend}, \& {Timmes}}]{paxton:18}
{Paxton}, B., {Schwab}, J., {Bauer}, E.~B., {et~al.} 2018,
  \bibinfo{title}{{Modules for Experiments in Stellar Astrophysics (MESA):
  Convective Boundaries, Element Diffusion, and Massive Star Explosions},}
  \apjs, 234, 34, \dodoi{10.3847/1538-4365/aaa5a8}

\bibitem[{B. {Paxton} {et~al.}(2019){Paxton}, {Smolec}, {Schwab}, {Gautschy},
  {Bildsten}, {Cantiello}, {Dotter}, {Farmer}, {Goldberg}, {Jermyn}, {Kanbur},
  {Marchant}, {Thoul}, {Townsend}, {Wolf}, {Zhang}, \& {Timmes}}]{paxton:19}
{Paxton}, B., {Smolec}, R., {Schwab}, J., {et~al.} 2019,
  \bibinfo{title}{{Modules for Experiments in Stellar Astrophysics (MESA):
  Pulsating Variable Stars, Rotation, Convective Boundaries, and Energy
  Conservation},} \apjs, 243, 10, \dodoi{10.3847/1538-4365/ab2241}

\bibitem[{P. {Podsiadlowski} {et~al.}(1995){Podsiadlowski}, {Cannon}, \&
  {Rees}}]{1995MNRAS.274..485P}
{Podsiadlowski}, P., {Cannon}, R.~C., \& {Rees}, M.~J. 1995,
  \bibinfo{title}{{The evolution and final fate of massive Thorne-Zytkow
  objects},} \mnras, 274, 485, \dodoi{10.1093/mnras/274.2.485}

\bibitem[{D.~J. {Price} \& J.~J. {Monaghan}(2007){Price} \&
  {Monaghan}}]{2007MNRAS.374.1347P}
{Price}, D.~J., \& {Monaghan}, J.~J. 2007, \bibinfo{title}{{An
  energy-conserving formalism for adaptive gravitational force softening in
  smoothed particle hydrodynamics and N-body codes},} \mnras, 374, 1347,
  \dodoi{10.1111/j.1365-2966.2006.11241.x}

\bibitem[{D. {Sanyal} {et~al.}(2015){Sanyal}, {Grassitelli}, {Langer}, \&
  {Bestenlehner}}]{Sanyal2015A&A...580A..20S}
{Sanyal}, D., {Grassitelli}, L., {Langer}, N., \& {Bestenlehner}, J.~M. 2015,
  \bibinfo{title}{{Massive main-sequence stars evolving at the Eddington
  limit},} \aap, 580, A20, \dodoi{10.1051/0004-6361/201525945}

\bibitem[{C. {Sch{\"u}rmann} {et~al.}(2025){Sch{\"u}rmann}, {Xu}, {Langer},
  {Lennon}, {Kruckow}, {Antoniadis}, {Haberl}, {Herrero}, {Kramer},
  {Schootemeijer}, {Shenar}, {Tauris}, \&
  {Wang}}]{Schurmann2025arXiv250323878S}
{Sch{\"u}rmann}, C., {Xu}, X.~T., {Langer}, N., {et~al.} 2025,
  \bibinfo{title}{{Populations of evolved massive binary stars in the Small
  Magellanic Cloud II: Predictions from rapid binary evolution},} \aap, 704,
  A219, \dodoi{10.1051/0004-6361/202554874}

\bibitem[{J. {Schwab} {et~al.}(2016){Schwab}, {Quataert}, \&
  {Kasen}}]{2016MNRAS.463.3461S}
{Schwab}, J., {Quataert}, E., \& {Kasen}, D. 2016, \bibinfo{title}{{The
  evolution and fate of super-Chandrasekhar mass white dwarf merger remnants},}
  \mnras, 463, 3461, \dodoi{10.1093/mnras/stw2249}

\bibitem[{Y. {Shao} \& X.-D. {Li}(2014){Shao} \&
  {Li}}]{Shao2014ApJ...796...37S}
{Shao}, Y., \& {Li}, X.-D. 2014, \bibinfo{title}{{On the Formation of Be Stars
  through Binary Interaction},} \apj, 796, 37,
  \dodoi{10.1088/0004-637X/796/1/37}

\bibitem[{Y. Shi {et~al.}(2026)Shi, Hu, Xu, Lin, \& Murray}]{Shi:2026kci}
Shi, Y., Hu, Q., Xu, Z., Lin, D. N.~C., \& Murray, N. 2026,
  \bibinfo{title}{Formation of black hole stars via star--black hole
  collisions,} arXiv e-prints.
\newblock \doarXiv{2608.27596}

\bibitem[{H.~M. {Tabernero} {et~al.}(2021){Tabernero}, {Dorda}, {Negueruela},
  \& {Marfil}}]{2021A&A...646A..98T}
{Tabernero}, H.~M., {Dorda}, R., {Negueruela}, I., \& {Marfil}, E. 2021,
  \bibinfo{title}{{The nature of VX Sagitarii. Is it a T{\.Z}O, a RSG, or a
  high-mass AGB star?},} \aap, 646, A98, \dodoi{10.1051/0004-6361/202039236}

\bibitem[{K.~S. {Thorne} \& A.~N. {Zytkow}(1975){Thorne} \&
  {Zytkow}}]{1975ApJ...199L..19T}
{Thorne}, K.~S., \& {Zytkow}, A.~N. 1975, \bibinfo{title}{{Red giants and
  supergiants with degenerate neutron cores.},} \apjl, 199, L19,
  \dodoi{10.1086/181839}

\bibitem[{K.~S. {Thorne} \& A.~N. {Zytkow}(1977){Thorne} \&
  {Zytkow}}]{1977ApJ...212..832T}
{Thorne}, K.~S., \& {Zytkow}, A.~N. 1977, \bibinfo{title}{{Stars with
  degenerate neutron cores. I. Structure of equilibrium models.},} \apj, 212,
  832, \dodoi{10.1086/155109}

\bibitem[{C.~A. {Tout} {et~al.}(2014){Tout}, {Zytkow}, {Church}, {Lau},
  {Doherty}, \& {Izzard}}]{2014MNRAS.445L..36T}
{Tout}, C.~A., {Zytkow}, A.~N., {Church}, R.~P., {et~al.} 2014,
  \bibinfo{title}{{HV2112, a Thorne-Zytkow object or a super asymptotic giant
  branch star.},} \mnras, 445, L36, \dodoi{10.1093/mnrasl/slu131}

\bibitem[{R.~H.~D. Townsend \& S. Teitler(2013)Townsend \&
  Teitler}]{TownsendTeitler2013}
Townsend, R. H.~D., \& Teitler, S. 2013, \bibinfo{title}{{GYRE}: An open-source
  stellar oscillation code based on a new {Magnus Multiple Shooting} scheme,}
  Monthly Notices of the Royal Astronomical Society, 435, 3406,
  \dodoi{10.1093/mnras/stt1533}

\bibitem[{C. {Wang} {et~al.}(2020){Wang}, {Langer}, {Schootemeijer}, {Castro},
  {Adscheid}, {Marchant}, \& {Hastings}}]{Wang2020ApJ...888L..12W}
{Wang}, C., {Langer}, N., {Schootemeijer}, A., {et~al.} 2020,
  \bibinfo{title}{{Effects of Close Binary Evolution on the Main-sequence
  Morphology of Young Star Clusters},} \apjl, 888, L12,
  \dodoi{10.3847/2041-8213/ab6171}

\bibitem[{L.~E. {Williams} {et~al.}(2025){Williams}, {Chang}, {Levesque}, \&
  {Quinn}}]{2025ApJ...993...61W}
{Williams}, L.~E., {Chang}, P., {Levesque}, E.~M., \& {Quinn}, T.~R. 2025,
  \bibinfo{title}{{Neutron Star─Main Sequence Collisions Robustly Form
  Dynamically Stable Thorne─{\.Z}ytkow Objects},} \apj, 993, 61,
  \dodoi{10.3847/1538-4357/ae100d}

\bibitem[{X.-T. {Xu} {et~al.}(2025){Xu}, {Sch{\"u}rmann}, {Langer}, {Wang},
  {Schootemeijer}, {Shenar}, {Ercolino}, {Haberl}, {Hastings}, {Jin}, {Kramer},
  {Lennon}, {Marchant}, {Sen}, {Tauris}, \& {de Mink}}]{Xu2025A&A...704A.218X}
{Xu}, X.-T., {Sch{\"u}rmann}, C., {Langer}, N., {et~al.} 2025,
  \bibinfo{title}{{Populations of evolved massive binary stars in the Small
  Magellanic Cloud: I. Predictions from detailed evolution models},} \aap, 704,
  A218, \dodoi{10.1051/0004-6361/202554786}

\end{thebibliography}

\end{CJK*}

\end{document}